\documentclass[journal]{IEEEtran}

\ifCLASSINFOpdf
\else
\fi
\usepackage{amsmath,amssymb}
\usepackage{algorithm}  
\usepackage{algorithmic}
\usepackage{graphicx,color,xcolor,rotating}

\DeclareMathOperator*{\argmin}{arg\,min}
\usepackage{cite}
\usepackage{graphicx}
\usepackage{float}
\usepackage{subfigure}
\usepackage{bm}
\usepackage{framed}
\usepackage{makecell}
\usepackage{threeparttable}

\newcommand{\bra}{{\mathbf a}}

\newcommand{\brn}{{\mathbf n}}

\newcommand{\brs}{{\mathbf s}}

\newcommand{\brw}{{\mathbf w}}
\newcommand{\brx}{{\mathbf x}}
\newcommand{\bry}{{\mathbf y}}

\newcommand{\brA}{{\mathbf A}}
\newcommand{\brB}{{\mathbf B}}
\newcommand{\brC}{{\mathbf C}}
\newcommand{\brD}{{\mathbf D}}
\newcommand{\brE}{{\mathbf E}}
\newcommand{\brF}{{\mathbf F}}
\newcommand{\brG}{{\mathbf G}}
\newcommand{\brH}{{\mathbf H}}
\newcommand{\brI}{{\mathbf I}}
\newcommand{\brJ}{{\mathbf J}}

\newcommand{\brL}{{\mathbf L}}
\newcommand{\brM}{{\mathbf M}}

\newcommand{\brO}{{\mathbf O}}
\newcommand{\brP}{{\mathbf P}}

\newcommand{\brS}{{\mathbf S}}

\newcommand{\brU}{{\mathbf U}}
\newcommand{\brV}{{\mathbf V}}
\newcommand{\brW}{{\mathbf W}}
\newcommand{\brX}{{\mathbf X}}
\newcommand{\brY}{{\mathbf Y}}
\newcommand{\brZ}{{\mathbf Z}}

\newcommand{\whatH}{\widehat{\mathbf H}}

\newcommand{\whatHk}{\widehat{\mathbf H}_k}
\newcommand{\wtildeHk}{\widetilde{\mathbf H}_k}

\newcommand{\Hk}{{\mathbf H}_k}

\newcommand{\Ek}{{\mathbf E}_k}

\newcommand{\Pc}{{\mathbf P}_c}
\newcommand{\Pp}{{\mathbf P}_p}
\newcommand{\Pk}{{\mathbf P}_k}
\newcommand{\Pj}{{\mathbf P}_j}
\newcommand{\Sc}{{\mathbf s}_c}
\newcommand{\Sk}{{\mathbf s}_k}

\newcommand{\tildePc}{\widetilde{\mathbf P}_c}
\newcommand{\tildePp}{\widetilde{\mathbf P}_p}

\newcommand{\In}{{\mathbf I}_N}
\newcommand{\sinr}{\text{SINR}}

\newcommand{\Lambdak}{\mathbf{\Lambda}_k}

\newcommand{\No}{\sigma_n^2}
\newcommand{\Ne}{\sigma_{e,k}^2}

\newcommand{\bbE}{\mathbb E}

\newcommand{\Rck}{R_{c,k}}
\newcommand{\Rpk}{R_{p,k}}

\newcommand{\Tr}{\text{tr}\slp}

\newcommand{\RD}{\mathbb{R}}
\newcommand{\CD}{\mathbb{C}}

\newcommand{\llp}{\left\{}
\newcommand{\lrp}{\right\}}
\newcommand{\slp}{\left(}
\newcommand{\srp}{\right)}
\newcommand{\sll}{\left|}
\newcommand{\srl}{\right|}
\newcommand{\mlp}{\left[}
\newcommand{\mrp}{\right]}
\newcommand{\diag}{\text{diag}}
\newcommand{\diagi}{\text{diag}^{-1}}
\newcommand{\blkdiag}{\text{blkdiag}}
\newcommand{\one}{{\mathbf 1}}
\newcommand{\brOmega}{\mathbf{\Omega}}
\newcommand{\brTheta}{\mathbf{\Theta}}

\newtheorem{theorem}{\textbf{Theorem}}
\newtheorem{lemma}{\textbf{Lemma}}

\newtheorem{remark}{Remark}

\begin{document}
\title{Robust Precoding Designs of RSMA for Multiuser MIMO Systems}

\author{Wentao~Zhou,~\IEEEmembership{Student Member,~IEEE,}
Yijie~Mao,~\IEEEmembership{Member,~IEEE,}
Di~Zhang,~\IEEEmembership{Senior Member,~IEEE,}\\
Mérouane Debbah,~\IEEEmembership{Fellow,~IEEE,}
and Inkyu Lee,~\IEEEmembership{Fellow,~IEEE}
\thanks{Wentao Zhou and Inkyu Lee are with the School of Electrical Engineering, Korea University, Seoul 02841, South Korea (e-mail: wtzhou@korea.ac.kr; inkyu@korea.ac.kr).}
\thanks{Yijie Mao is with the School of Information Science and Technology, ShanghaiTech University, Shanghai 201210, China (e-mail: maoyj@shanghaitech.edu.cn).}
\thanks{Di Zhang is with the School of Electrical and Information Engineering, Zhengzhou University, Zhengzhou 450001, China (e-mail: dr.di.zhang@ieee.org).}
\thanks{M\'{e}rouane Debbah is with the Department of Electrical Engineering and Computer Science, Khalifa University of Science and Technology, Abu Dhabi, United Arab Emirates (e-mail: merouane.debbah@ku.ac.ae).}
\thanks{This work has been accepted for publication in IEEE Transactions on Wireless Communications. Personal use of this material is permitted. Permission from IEEE must be obtained for all other uses, in any current or future media, including reprinting/republishing this material for advertising or promotional purposes, creating new collective works, for resale or redistribution to servers or lists, or reuse of any copyrighted component of this work in other works.}
}

\maketitle

\begin{abstract}
Rate-splitting multiple access (RSMA) has been studied for multiuser multiple-input multiple-output (MU-MIMO) systems especially in the presence of imperfect channel state information (CSI) at the transmitter.
However, its precoding designs that maximize the sum rate normally have high computational complexity.
To implement an efficient RSMA scheme for the MU-MIMO system, in this work, we propose a novel robust precoding design, which can handle imperfect CSI.
Specifically, we first adopt the generalized mutual information to construct a lower bound of the objective function in the sum rate maximization problem.
Then, we apply a smooth lower bound of the non-smooth sum rate objective function to construct a new optimization problem.
By revealing the relationship between the generalized signal-to-interference-plus-noise ratio and the minimum mean square error matrices, we transform the constructed problem into a tractable one.
After decomposing the transformed problem into three subproblems, we investigate a new alternating precoding design based on sequential solutions.
Simulation results demonstrate that the proposed precoding scheme achieves comparable performance to conventional methods, while significantly reducing the computational complexity.
\end{abstract}


\IEEEpeerreviewmaketitle

\section{Introduction}
Precoding techniques have played a key role in multiple-input multiple-output (MIMO) communications to maximize the sum rate \cite{GPIP20}.
Effective inter-user interference mitigation or sum rate maximization always requires perfect channel state information (CSI) at the transmitter (CSIT).
However, practical limitations such as channel estimation errors \cite{ESLSMA13, RDBPS19, GPIP20} and quantization errors \cite{LFBD08, CQIA12, RP23} inevitably lead to inaccurate CSIT.
Failing to incorporate CSIT errors in the precoding design results in severe sum rate degradations, particularly in high signal-to-noise ratio (SNR) regions. 
Therefore, it is crucial to develop robust precoding schemes which can address the impact of CSIT imperfection.

Rate-splitting multiple access (RSMA) has been proposed as a viable downlink transmission strategy for multiuser MIMO (MU-MIMO) systems \cite{SRRS16, RTRS16, RSMIMO16, RSWMMSE22, RSMAto23}.
Owing to its high degree of freedom, RSMA has been widely studied in imperfect CSIT and overloaded scenarios.
The essence of the RSMA lies in the common signal construction at a base station (BS) and its removal at each user through successive interference cancellation (SIC).
From an information-theoretic perspective, RSMA can be regarded as a generalization of the Han-Kobayashi scheme \cite{HK81}.
Existing works have shown that in the presence of imperfect CSIT, RSMA achieves explicit performance gains over linearly-precoded space-division multiple access (SDMA), orthogonal multiple access (OMA), non-orthogonal multiple access (NOMA), and non-linearly precoded dirty paper coding (DPC) \cite{RSMA19, RSDPC20, RSMAfu22}.
Furthermore, deploying RSMA structures has proven effective in enhancing system performance in various applications such as cloud radio access network (C-RAN), integrated sensing and communication (ISAC), satellite communication, and reconfigurable intelligent surfaces (RIS) \cite{RSMACRAN19, RSMAISAC21, RSMAURLLC21, RSsate21, RSMARIS22, RSMAUP23}.
Careful determination of precoding matrices is crucial in RSMA systems \cite{RSmMIMO16, RSSOS23}.
However, due to the non-smooth function within the achievable common rate, maximizing the sum rate in RSMA brings nontrivial challenges in the precoding design.

There have been some efforts for the precoding design in RSMA.
A precoding design which eliminates inter-group interference was proposed by utilizing a sum rate upper bound \cite{RSMAIN23}.
Based on the regularized block diagonalization (RBD) scheme \cite{BD04, GCI09}, a linear precoding and stream combining technique was investigated in \cite{LPSC20}.
Also, a successive null-space (SNS) precoding scheme was studied in \cite{DLSNS22}.
To overcome the limitation on the number of common streams, a generalized singular value decomposition (GSVD) based precoding method was proposed in \cite{GSVD22}.
However, all the aforementioned precoding approaches failed to maximize the sum rate when dealing with imperfect CSIT.

To maximize the ergodic sum rate (ESR) of RSMA, which reflects the sum rate performance of all possible channel realizations, researchers have employed various optimization techniques \cite{SRRS16, RTRS16, RSWMMSE22, CCCP20, OPBFRSMA23, LSE23}.
The sample average approximation (SAA) based weighted minimum mean square error (WMMSE) precoding scheme \cite{SRRS16} transformed the stochastic non-convex optimization problem into a series of quadratically constrained quadratic programs (QCQP), which can be solved by using the interior-point method iteratively.
This approach has also been extended to the robust max-min fairness precoding design \cite{RTRS16} and the robust precoding design for maximizing the ESR in MU-MIMO systems \cite{RSWMMSE22}.
Besides, the concave-convex procedure (CCP) based linear precoding scheme \cite{CCCP20} changed the original optimization problem into a series of successively refined convex problems.
However, the above methods require high computational complexity to solve the optimization problem in each iteration.
To reduce the computing task, a fractional programming based precoding scheme was proposed in \cite{OPBFRSMA23} assuming perfect CSIT.
Considering imperfect CSIT, the authors in \cite{LSE23} approximated the minimum common rate and presented a precoding scheme that iteratively finds the eigenvector to maximize the ESR.
Nevertheless, these schemes are designed for single-receive antenna systems, thereby they are not applicable to the considered MU-MIMO system.

In this paper, we study a novel robust precoding design which maximizes the ESR in RSMA systems with imperfect CSIT while maintaining low complexity.
The main contributions of this paper are summarized as follows:
\begin{itemize}
\item When only imperfect CSIT is available, we first derive a lower bound of the ESR by examining the generalized mutual information.
We then reformulate the derived optimization problem by applying a lower bound of the minimum function.
To simplify the expression, we reveal the relationship between the generalized signal-to-interference-plus-noise ratio (SINR) and the minimum mean square error (MMSE) matrices.
To the best of our knowledge, this relationship has not been investigated before.
Based on this derivation, we finally transform the reformulated optimization problem into a tractable one.

\item To solve the transformed problem, we decompose it into three subproblems by employing the block coordinate descent (BCD) approach \cite{NP99}.
Leveraging the sequential solutions of these subproblems, we propose a new alternating precoding design.
Compared to the state-of-the-art WMMSE-SAA scheme \cite{RSWMMSE22}, which relies on the generation of CSIT samples and an optimization toolbox, the proposed scheme can reduce the computational complexity.
Simulation results demonstrate that the proposed scheme achieves comparable performance to the WMMSE-SAA scheme with lower complexity.
\end{itemize}

The remainder of this paper is structured as follows:
Section \ref{SystemModel} introduces the system model.
The problem formulation is given in Section \ref{ProbReform}.
Section \ref{ProposedSchemes} elaborates on the proposed precoding design.
Simulation results are illustrated in Section \ref{SimuResults}, and Section \ref{Conclusion} concludes this paper.

Notation: Column vectors and matrices are denoted by lower-case bold and upper-case bold letters, respectively.
$\slp \cdot \srp^T$, $\slp \cdot \srp^*$, $\slp \cdot \srp^H$, $\sll \cdot \srl$, and $\Tr\cdot\srp$ stand for transpose, conjugate, Hermitian, determinant, and trace, respectively.
$\diag\llp \bra \lrp$ represents a diagonal matrix whose diagonal elements consist of the elements of $\bra$, and $\diagi \llp \brA \lrp$ collects diagonal elements of $\brA$ into a column vector.
$\blkdiag\llp \cdot \lrp$ indicates a block diagonal matrix.
$\In$ equals an $N\times N$ identity matrix, while $\one_{M\times N}$ refers to an $M\times N$ all-ones matrix.
$\bbE\mlp \cdot \mrp$ is the statistical expectation operation.
We define the complex partial derivative operation $\frac{\partial f\slp\brA \srp}{\partial \brA^*}$ as $\nabla_{\brA} f$.

\section{System Model}
\label{SystemModel}
We consider a single-cell narrowband MU-MIMO system with $K$ users, where a BS is equipped with $M$ transmit antennas and each user has $N$ receive antennas.
Without loss of generality, it is assumed that $M>N$.
The block fading channel between the BS and user $k$ is denoted as $\Hk \in \CD^{M\times N}$, whose elements are zero-mean independent and identically distributed (i.i.d.) unit variance complex Gaussian random variables.
The received signal at user $k$ is expressed as
\begin{equation*}
\begin{aligned}
\bry_k = \Hk^H \brx + \brn_k,
\end{aligned}
\end{equation*}
where $\brx \in \CD^{M\times 1}$ denotes the transmit signal and $\brn_k\in \CD^{N\times 1}$ is an i.i.d. complex Gaussian noise vector with zero mean and variance $\sigma_n^2$.
A total power constraint is imposed as $\bbE \mlp \brx^H \brx \mrp = \rho$, where $\rho$ is the maximum transmit power.

\subsection{RSMA for MU-MIMO}
\label{RSMIMO}

\begin{figure}[t]
\centering
\subfigure[BS structure.]{
\label{fig_RSMA_BS}
\includegraphics[width=2.5in]{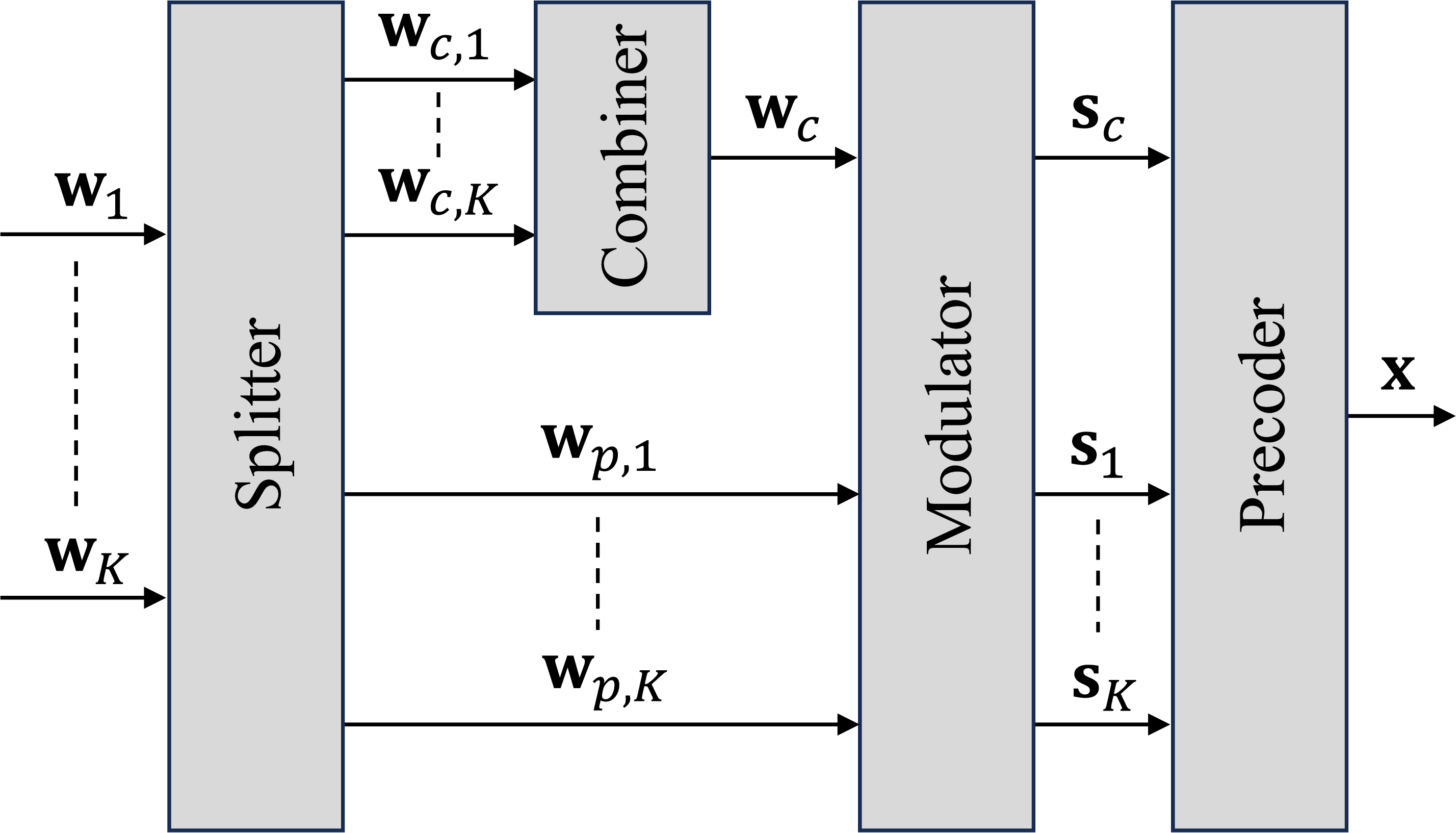}}
\subfigure[Receiver structure of the $k$-th user.]{
\label{fig_RSMA_UE}
\includegraphics[width=2.5in]{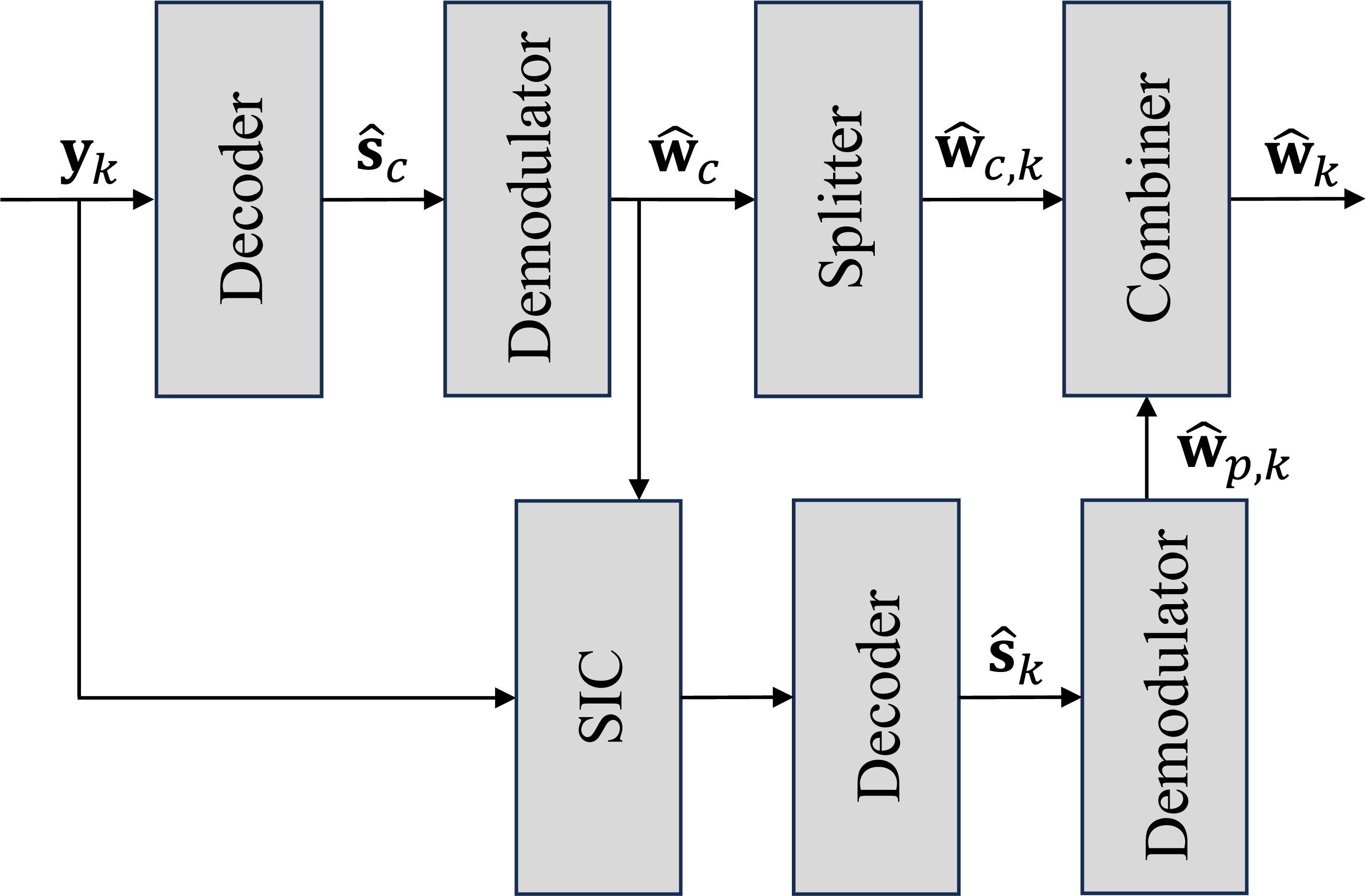}}
\caption{System model of 1-layer RSMA for MU-MIMO systems.}
\label{fig_RSMA_structure}
\end{figure}

As illustrated in Fig. \ref{fig_RSMA_structure}, in this paper, we focus on a 1-layer RSMA for MU-MIMO systems, where SIC is applied once \cite{RSWMMSE22, RSMAto23}.
The data streams $\llp \brw_k \lrp$ are split into common and private parts.
The common parts $\llp \brw_{c,k} \lrp$ are combined into $N$ common messages $\brw_c$ using a predefined codebook\footnote{It was shown in \cite{RSWMMSE22} that the system performance is maximized when the number of common messages equals that of receive antennas.}.
Then, the common and private messages are modulated, linearly precoded, and finally superimposed.
This process yields the transmit signal $\brx$ as
\begin{equation*}
\begin{aligned}
\brx = \Pc \Sc + \sum_{k=1}^K \Pk \Sk, 
\end{aligned}
\end{equation*}
where $\Pc \in \CD^{M\times N}$ and $\Pk \in \CD^{M\times N}$ are the precoding matrices for the common symbol vector $\Sc \in \CD^{N\times 1}$ and private symbol vector $\Sk \in \CD^{N\times 1}$ for the $k$-th user, respectively.
It is assumed that both $\Sc$ and $\Sk$ are uncorrelated with $\bbE\mlp \Sc\Sc^H \mrp = \In$, $\bbE\mlp \Sk\Sk^H \mrp = \In$, and $\bbE\mlp \Sc\Sk^H \mrp = \mathbf{0}$.
Thus, the power constraint can be rewritten as $\Tr \Pc\Pc^H\srp + \sum_{k=1}^K\Tr \Pk\Pk^H\srp = \rho$.
If $\Tr \Pc\Pc^H\srp = 0$, there is no common signal and the system boils down to the SDMA structure \cite{RSWMMSE22, RSMAto23}. 

After receiving $\bry_k$, each user first decodes the common symbol vector $\Sc$ by treating all private signals as interference.
The decoded common signal is then removed from $\bry_k$ through the SIC as $\bry_k - \Hk^H \Pc \Sc$.
Subsequently, each user decodes its own private symbol vector by treating other private signals as interference.
The instantaneous common and private rates for user $k$ respectively represent
\begin{equation*}
\begin{aligned}
\Rck = \log_2 \sll \In + \sinr_{c,k} \srl,
\end{aligned}
\end{equation*}
\begin{equation*}
\begin{aligned}
\Rpk = \log_2 \sll \In + \sinr_{p,k} \srl,
\end{aligned}
\end{equation*}
where the SINR matrices for common and private signals are respectively written as
\begin{equation*}
\begin{aligned}
\sinr_{c,k} \! = \! \Pc^H \Hk \slp \sum\nolimits_{j=1}^{K} \Hk^H \Pj \Pj^H \Hk + \No \In \srp^{-1} \! \Hk^H \Pc,
\end{aligned}
\end{equation*}
\begin{equation*}
\begin{aligned}
\sinr_{p,k} \! = \! \Pk^H \Hk \slp \sum\nolimits_{j\neq k} \Hk^H \Pj \Pj^H \Hk + \No \In \srp^{-1}\Hk^H \Pk.
\end{aligned}
\end{equation*}
From Fig. \ref{fig_RSMA_BS}, it can be seen that the common message $\brw_c$ consists of a part of the information from all users.
In order to successfully perform the SIC at each user, the achievable common rate of RSMA should be the minimum common rate across all users.
Thus, the instantaneous sum rate is computed as
\begin{equation*}
\begin{aligned}
R = \min_{k=1,\cdots,K} \llp\Rck\lrp + \sum_{k=1}^K \Rpk.
\end{aligned}
\end{equation*}

\subsection{Imperfect CSIT Model}
\label{ImpCSIT}
Due to practical limitations, CSIT is always imperfect.
In the following, we introduce imperfect CSIT models with channel estimation and quantization errors.

\subsubsection{Channel estimation error model}
\label{ImpCSITesti}
When the channel follows a complex Gaussian distribution and the BS adopts a linear MMSE estimator, the actual channel $\Hk$ is composed of the estimated channel $\whatHk$ and its associated error $\Ek$ which are independent of each other as \cite{CE06, LCRD22}
\begin{equation}
\label{CSITesti}
\begin{aligned}
\Hk = \whatHk + \Ek,
\end{aligned}
\end{equation}
where $\whatHk$ and $\Ek$ are zero-mean complex Gaussian matrices comprised of i.i.d. elements with variance $1-\Ne$ and $\Ne$, respectively.
The parameter $\Ne$ serves as an indicator of the estimation error level.

\subsubsection{Channel quantization error model}
\label{ImpCSITquan}
When each user employs limited feedback to convey the downlink CSI, the CSIT error mainly arises from the quantization process \cite{DLLF20, DLLF22}.
Consider a rank-$N$ eigenvalue decomposition $\Hk\Hk^H=\wtildeHk \Lambdak \wtildeHk^H$, where $\wtildeHk \in \CD^{M \times N}$ consists of the orthonormal eigenvectors of $\Hk\Hk^H$ corresponding to its $N$ non-zero eigenvalues, and $\Lambdak \in \CD^{N \times N}$ is a diagonal matrix whose diagonal elements are the non-zero eigenvalues.
The quantization process is achieved by finding the minimum chordal distance between $\wtildeHk$ and semi-unitary matrices from a predefined codebook with $2^B$ codewords.
For instance, user $k$ selects the $i$-th codeword $\brC_{k,i}$ from codebook $\mathcal{C}_{k}=\llp \brC_{k,1},\cdots,\brC_{k,2^B} \lrp$ based on the following criterion as
\begin{equation*}
\begin{aligned}
\whatHk = \argmin_{\brC_{k,i} \in \mathcal{C}_k} d^2 \slp \brH_k,\brC_{k,i} \srp,
\end{aligned}
\end{equation*}
where $d^2 \slp \brH_k,\brC_{k,i} \srp$ denotes the chordal distance $N-\text{tr}\slp \wtildeHk^H \brC_{k,i} \brC_{k,i}^H \wtildeHk \srp$.

Then, $\wtildeHk$ can be decomposed as \cite{LFBD08}
\begin{equation*}
\begin{aligned}
\wtildeHk = \whatHk \brX_k \brY_k + \brS_k \brZ_k,
\end{aligned}
\end{equation*}
where $\brX_k$ represents a unitary matrix, $\brY_k$ and $\brZ_k$ stand for $N\times N$ upper triangular matrices with positive diagonal elements satisfying $\brY_k^H \brY_k + \brZ_k^H\brZ_k = \In$ and $\Tr \brZ_k^H\brZ_k \srp = d^2 \slp \brH_k,\whatHk \srp$, and the semi-unitary matrix $\brS_k \in \CD^{M\times N}$ is isotropically distributed in an $N$-dimensional plane within the $\slp M-N \srp$-dimensional left nullspace of $\whatHk$.
$\whatHk$, $\brX_k$, and $\brY_k$ are independent of each other, and $\brS_k$ and $\brZ_k$ are also independent.

Finding a robust precoding design based on the above channel decomposition is nontrivial since only quantized channels are available at the BS \cite{JALFP22}.
Thus, the channel $\Hk$ is approximated by \cite{RP23}
\begin{equation}
\begin{aligned}
\overline{\brH}_k = \delta _k \whatHk + \eta_k \brO_k,
\end{aligned}
\end{equation}
where $\delta_k$ and $\eta_k$ represent random values with $\bbE\mlp \delta_k\mrp = \sqrt{M - \frac{M^2 \gamma}{M-N}}$ and $\bbE\mlp \eta_k\mrp = \sqrt{\frac{M^2\gamma}{M-N}}$, and $\brO_k$ indicates a random semi-unitary matrix independent of $\whatHk$.
The parameter $\gamma=\frac{1}{N}\bbE\llp \min_{\brC_{k,i} \in \mathcal{C}_k} d^2 \slp \brH_k,\brC_{k,i} \srp \lrp$ denotes the quantization distortion for each column of $\Hk$.

Due to the presence of CSIT errors, the precoding design introduces additional interference if we treat $\whatHk$ as the actual channel.
Therefore, instead of the instantaneous sum rate, we maximize the ESR as
\begin{subequations}
\label{asr_prob}
\begin{align}
\label{asr_prob_obj}
&\max_\brP \quad \min_{k=1,\cdots,K} \llp \bbE_{\whatHk,\Ek} \mlp R_{c,k} \mrp \lrp + \sum_{k=1}^K \bbE_{\whatHk,\Ek} \mlp R_{p,k} \mrp \\
\label{asr_prob_con}
&\text{subject to} \quad \Tr \brP\brP^H \srp = \rho,
\end{align}
\end{subequations}
where $\brP = \mlp  \Pc, \brP_1,\cdots,\brP_K \mrp \in \CD^{M\times N \slp K+1 \srp}$ equals the concatenation of precoding matrices.
We can check that the above problem is hard to solve due to its non-smooth objective function.

\section{Problem Formulation}
\label{ProbReform}
In this section, we focus only on the channel estimation error model \eqref{CSITesti} in Section \ref{ImpCSIT}, as an extension to the channel quantization error model is trivial.
To facilitate the derivation, we transform the optimization problem \eqref{asr_prob} with lower bounds.
In the following, we first investigate a lower bound of the objective in \eqref{asr_prob_obj}.
Rewriting the received signal at user $k$ in terms of the channel estimation error model yields
\begin{equation*}
\begin{aligned}
\bry_k = \ & \whatHk^H \Pc \Sc +  \whatHk^H \sum_{k=1}^K \Pk \Sk \\
& + \Ek^H \Pc \Sc +  \Ek^H \sum_{k=1}^K \Pk \Sk + \brn_k.
\end{aligned}
\end{equation*}

Based on the received signal, it can be seen that calculating the common and private rates becomes challenging due to the unknown CSIT error $\Ek$.
To address this, we adopt generalized mutual information, which is defined as the worst-case mutual information (see \cite{GMI10}, \cite{mmfairRSMA23} and references therein).
In this model, the terms associated with the unknown CSIT error are treated as the independent Gaussian noise.
Then we can obtain the generalized SINR matrices for the common and private signals respectively as
\begin{equation*}
\begin{aligned}
\overline{\sinr}_{c,k} = \Pc^H \whatHk \slp \brJ_{c,k} + \brL_{c,k} + \No\In \srp^{-1}\whatHk^H \Pc,
\end{aligned}
\end{equation*}
\begin{equation*}
\begin{aligned}
\overline{\sinr}_{p,k} =  \Pk^H \whatHk \slp \brJ_{p,k} + \brL_{p,k} + \No\In \srp^{-1}\whatHk^H \Pk,
\end{aligned}
\end{equation*}
where $\brJ_{c,k}= \whatHk^H\Pp\Pp^H\whatHk$ with $\Pp=\mlp \brP_1,\cdots,\brP_K \mrp \in \CD^{M\times NK}$ being the concatenation of private precoding matrices, $\brL_{c,k} = \Ek^H\brP\brP^H\Ek$, $\brJ_{p,k} = \sum_{j\neq k} \whatHk^H\Pj\Pj^H\whatHk$, and $\brL_{p,k} = \Ek^H\brP_p\brP_p^H\Ek$.

Based on the generalized SINR matrices, we have $\overline{R}_{z,k} \triangleq \log_2 \sll \In + \overline{\sinr}_{z,k} \srl \leq R_{z,k}$ and $\bbE_{\Hk}\mlp \overline{R}_{z,k} \mrp \leq \bbE_{\Hk}\mlp R_{z,k}\mrp$ for $z\in\llp c,p \lrp$.
Introducing Jensen's inequality to the above expressions further gives \cite{logdet_tit01}
\begin{equation}
\label{rate_com_lb}
\begin{aligned}
\bbE_{\Hk}\mlp \overline{R}_{z,k} \mrp & = \bbE_{\whatHk}\mlp \bbE_{\Ek|\whatHk}\mlp \log_2 \sll \In + \overline{\sinr}_{z,k} \srl \mrp\mrp \\
& \geq \bbE_{\whatHk}\mlp \log_2 \sll \In + \widehat{\sinr}_{z,k} \srl \mrp, \\
\end{aligned}
\end{equation}
where $\widehat{\sinr}_{z,k}$ for $z\in\llp c,p \lrp$ represents the conditional expectation based generalized SINR matrices for the common and private signals respectively as
\begin{equation}
\label{sinr_com_lbe}
\begin{aligned}
\widehat{\sinr}_{c,k} \! \triangleq \! \Pc^H \whatHk \slp \brJ_{c,k} \! + \! \bbE_{\Ek} \! \mlp \brL_{c,k} \mrp + \No\In \srp^{-1}\whatHk^H \Pc,
\end{aligned}
\end{equation}
\begin{equation}
\label{sinr_pri_lbe}
\begin{aligned}
\widehat{\sinr}_{p,k} \! \triangleq \! \Pk^H \whatHk \slp \brJ_{p,k} \! + \! \bbE_{\Ek} \! \mlp \brL_{p,k} \mrp \! + \No\In \srp^{-1}\whatHk^H \Pk.
\end{aligned}
\end{equation}
It is worth noting that the original optimization problem \eqref{asr_prob} aims to maximize the ergodic performance under the assumption of perfect CSIT, whereas the expression in \eqref{rate_com_lb} accounts for CSIT uncertainty by conditioning on the imperfect CSIT $\whatHk$ and considering all possible realizations of $\Ek$.
Consequently, the optimization problem based on this expression enables a robust solution that explicitly incorporates the impact of CSIT errors given $\whatHk$.
Observing \eqref{sinr_com_lbe} and \eqref{sinr_pri_lbe}, we notice the presence of two undetermined terms associated with the expectation of $\Ek$.
To clarify these terms, we introduce the following lemma.
\begin{lemma} [{\cite[Lemma 4]{RP23}}]
\label{lemma_expectation}
Let us denote $\brX \in \CD^{M \times M}$ as a fixed matrix and $\brY \in \CD^{M \times N}$ as a random matrix.
The $\slp m,n \srp$-th element of $\brY$ is i.i.d. with zero mean and variance $\sigma_{mn}^2$.
Then, we have
\begin{equation*}
\begin{aligned}
\bbE \mlp \brY^H \brX \brY \mrp = \text{diag}\llp \mathbf{\Theta} \text{diag}^{-1}\llp \brX \lrp \lrp,
\end{aligned}
\end{equation*}
where the $\slp n,m \srp$-th element of $\mathbf{\Theta} \in \RD^{N \times M}$ equals $\sigma_{mn}^2$.
$\hfill \blacksquare$
\end{lemma}

From this lemma, if all the elements of $\brY$ have the same variance $\sigma^2$, we further have
\begin{equation*}
\begin{aligned}
\bbE \mlp \brY^H \brX \brY \mrp = \sigma^2 \Tr \brX \srp \brI_N.
\end{aligned}
\end{equation*}
Then, the undetermined terms in \eqref{sinr_com_lbe} and \eqref{sinr_pri_lbe} can be rewritten respectively as
\begin{equation}
\label{sinr_com_lb}
\begin{aligned}
\bbE_{\Ek} \mlp \brL_{c,k} \mrp = \Ne \Tr \brP \brP^H \srp \brI_N,
\end{aligned}
\end{equation}
\begin{equation}
\label{sinr_pri_lb}
\begin{aligned}
\bbE_{\Ek} \mlp \brL_{p,k} \mrp = \Ne \Tr \Pp\Pp^H \srp \brI_N.
\end{aligned}
\end{equation}
Now, we have explicit expressions of $\widehat{\sinr}_{z,k}$ for $z\in\llp c,p\lrp$.

Originally, the expectation in \eqref{rate_com_lb} captures the ergodic performance over all possible partial CSITs.
In case when imperfect CSIT $\llp \whatHk \lrp$ is given, we can neglect the expectation operation.
Thus, we can optimize a lower bound of \eqref{asr_prob_obj} based on \eqref{rate_com_lb} as
\begin{subequations}
\label{asr_prob_lb}
\begin{align}
\label{asr_prob_lb_obj}
&\max_\brP \min_{k=1,\cdots,K} \! \llp \log \sll \In \! + \! \widehat{\sinr}_{c,k} \srl \lrp \! + \! \sum_{k=1}^K \log \sll \In \! + \! \widehat{\sinr}_{p,k} \srl \\
\label{asr_prob_lb_con}
&\text{subject to} \quad \Tr \brP\brP^H \srp = \rho,
\end{align}
\end{subequations}
where we delete the scaling factor $\log2$ since it does not affect the solution.
Compared with the WMMSE-SAA precoding scheme, the above optimization problem avoids generating CSIT samples to determine the objective function.
However, finding the optimal solution is still challenging due to the non-smooth objective function.
In what follows, we introduce a lower bound to the above problem.

To smooth the minimum function, we apply the log-sum-exp technique as \cite{CO04}
\begin{equation}
\label{LSE}
\begin{aligned}
\min_{k=1,\cdots,K} \llp x_k \lrp \geq - \log \slp \sum_{k=1}^K \exp \slp - x_k \srp \srp.
\end{aligned}
\end{equation}
By incorporating \eqref{LSE} into problem \eqref{asr_prob_lb}, we have
\begin{subequations}
\label{lse_prob_plb}
\begin{alignat}{2}
\label{lse_prob_plb_obj}
& \max_\brP \quad  &&  - \log \slp \sum_{k=1}^K \exp \slp - \log \sll \In + \widehat{\sinr}_{c,k} \srl \srp \srp \nonumber \\
&&& + \sum_{k=1}^K \log \sll \In + \widehat{\sinr}_{p,k} \srl \\
&\text{subject to} \quad && \Tr \brP\brP^H \srp = \rho.
\end{alignat}
\end{subequations}
Although \eqref{lse_prob_plb_obj} is smooth, problem \eqref{lse_prob_plb} is NP-hard.
In the following, we aim to solve this problem through the BCD approach.

\section{Proposed Low-complexity Robust Precoding Design}
\label{ProposedSchemes}
In this section, we first study an alternating algorithm to optimize the formulated problem \eqref{lse_prob_plb}.
Then, we analyze the convergence and computational complexity of the proposed algorithm.

\subsection{Proposed Precoding Design}
\label{EstiRWMMSE}
We first construct an equivalent problem of \eqref{lse_prob_plb}.
Let us denote $\brD_{c,k}\in\CD^{N\times N}$ and $\brD_{p,k}\in\CD^{N\times N}$ as the receive filters for the common and private signals of user $k$, respectively.
The estimates of the common and private symbol vectors are then calculated by $\widehat{\brs}_{c,k}=\brD_{c,k}\bry_k$ and $\widehat{\brs}_k=\brD_{p,k}\slp \bry_k - \Hk^H \Pc \Sc \srp$.
Since only imperfect CSIT is available, we investigate the conditional expectation based MSE matrices as
\begin{equation}
\label{cemse_com}
\begin{aligned}
{\brM}_{c,k} = &\ \bbE \mlp \slp \Sc- \widehat{\brs}_{c,k}\srp \slp \Sc- \widehat{\brs}_{c,k}\srp^H \mrp \\
 = &\ \In - \Pc^H \whatHk \brD_{c,k}^H - \brD_{c,k} \whatHk^H \Pc \\
 & + \brD_{c,k}\whatHk^H \brP\brP^H \whatHk\brD_{c,k}^H \\
 & + \bbE_{\Ek} \mlp \brD_{c,k} \brL_{c,k} \brD_{c,k}^H \mrp + \No \brD_{c,k}\brD_{c,k}^H,
\end{aligned}
\end{equation}
\begin{equation}
\label{cemse_pri}
\begin{aligned}
{\brM}_{p,k} = &\ \bbE \mlp \slp \Sk- \widehat{\brs}_{k}\srp \slp \Sk- \widehat{\brs}_{k}\srp^H \mrp \\
 = &\ \In - \Pk^H \whatHk \brD_{p,k}^H - \brD_{p,k} \whatHk^H \Pk \\
& + \brD_{p,k}\whatHk^H \brP_p\brP_p^H \whatHk\brD_{p,k}^H \\
& + \bbE_{\Ek} \mlp \brD_{p,k} \brL_{p,k} \brD_{p,k}^H \mrp + \No \brD_{p,k}\brD_{p,k}^H.
\end{aligned}
\end{equation}

Then, we obtain the optimum receive filters and MMSE matrices for given precoding matrices.
Based on \eqref{cemse_com} and \eqref{cemse_pri}, the optimum receive filters are computed by solving $\nabla_{\brD_{c,k}}\Tr {\brM}_{c,k}\srp =\mathbf{0}$ and $\nabla_{\brD_{p,k}}\Tr{\brM}_{p,k}\srp =\mathbf{0}$ as
\begin{equation}
\label{cerf_com}
\begin{aligned}
\brD_{c,k}^\text{MMSE} = \Pc^H \whatHk \brF_k^{-1},
\end{aligned}
\end{equation}
\begin{equation}
\label{cerf_pri}
\begin{aligned}
\brD_{p,k}^\text{MMSE} =  \Pk^H \whatHk \brG_k^{-1},
\end{aligned}
\end{equation}
where $\brF_k = \whatHk^H \brP\brP^H \whatHk + \Ne \Tr \brP \brP^H \srp \brI_N + \No\In$ and $\brG_k = \whatHk^H \brP_p\brP_p^H \whatHk + \Ne \Tr \Pp\Pp^H \srp \brI_N + \No\In$.

One should note that the optimum receive filters in \eqref{cerf_com} and \eqref{cerf_pri} are not the actual MMSE receive filters.
Since only imperfect CSI is available, the obtained receive filters are utilized to facilitate a robust precoding design at the BS.
Substituting \eqref{cerf_com} and \eqref{cerf_pri} into \eqref{cemse_com} and \eqref{cemse_pri}, respectively, the conditional expectation based MMSE matrices are written by
\begin{equation}
\label{cemmse_com}
\begin{aligned}
{\brM}_{c,k}^\text{MMSE} \triangleq \In - \Pc^H \whatHk \brF_k^{-1}\whatHk^H\Pc,
\end{aligned}
\end{equation}
\begin{equation}
\label{cemmse_pri}
\begin{aligned}
{\brM}_{p,k}^\text{MMSE} \triangleq \In - \Pk^H \whatHk \brG_k^{-1}\whatHk^H\Pk.
\end{aligned}
\end{equation}
Now, we try to reformulate problem \eqref{lse_prob_plb} through the above MMSE matrices by introducing the following lemma.
\begin{lemma}
\label{lemma_SINRandMMSE}
The generalized SINR and MMSE matrices based on the conditional expectation are expressed as
\begin{equation*}
\begin{aligned}
\slp{\brM}_{z,k}^\text{MMSE} \srp^{-1}= \In + \widehat{\sinr}_{z,k} \ \text{for} \ z\in \llp c,p\lrp.
\end{aligned}
\end{equation*}
\end{lemma}
\begin{IEEEproof}
Let us write $\brF_k$ and $\brG_k $ as
\begin{equation*}
\begin{aligned}
& \brF_k =  \whatHk^H \Pc\Pc^H \whatHk + \brJ_{c,k} + \Ne \Tr \brP \brP^H \srp \brI_N + \No\In, \\
& \brG_k = \whatHk^H\Pp\Pp^H\whatHk + \brJ_{p,k} + \Ne \Tr \Pp \Pp^H \srp \brI_N + \No\In.
\end{aligned}
\end{equation*}
Applying the matrix inversion lemma to ${\brM}_{c,k}^\text{MMSE}$, we have
\begin{equation*}
\begin{aligned}
{\brM}_{c,k}^\text{MMSE} = & \Big( \Pc^H\whatHk \slp \brJ_{c,k} + \slp \Ne \Tr \brP \brP^H \srp + \No\srp\In \srp ^{-1} \\
& \times \whatHk^H\Pc  + \In \Big)^{-1}\\
= & \slp \In + \widehat{\sinr}_{c,k} \srp^{-1}.
\end{aligned}
\end{equation*}
Similarly, ${\brM}_{p,k}^\text{MMSE}$ is written as
\begin{equation*}
\begin{aligned}
{\brM}_{p,k}^\text{MMSE} = & \Big( \Pk^H\whatHk \slp \brJ_{p,k} + \slp \Ne \Tr \Pp \Pp^H \srp + \No \srp \In \srp ^{-1}\\
& \times \whatHk^H\Pk + \In \Big)^{-1}\\
= & \slp \In + \widehat{\sinr}_{p,k} \srp^{-1}.
\end{aligned}
\end{equation*}
From the above derivations, we can obtain Lemma \ref{lemma_SINRandMMSE}.
This completes the proof.
\end{IEEEproof}

From Lemma \ref{lemma_SINRandMMSE}, we can see that \eqref{cemmse_com} and \eqref{cemmse_pri} can be rewritten by \eqref{sinr_com_lbe} and \eqref{sinr_pri_lbe}, respectively.
Consequently, problem \eqref{lse_prob_plb} is recast as
\begin{subequations}
\label{lse_prob_lb}
\begin{alignat}{2}
\label{lse_prob_lb_obj}
&\min_\brP \quad && f_1\slp \brP \srp \triangleq \log\slp \sum_{k=1}^K \exp \slp \log \sll  {\brM}_{c,k}^\text{MMSE} \srl \srp \srp \nonumber\\
&&& + \sum_{k=1}^K \log\sll {\brM}_{p,k}^\text{MMSE} \srl \\
\label{lse_prob_lb_con}
&\text{subject to} \quad && \Tr \brP\brP^H \srp = \rho.
\end{alignat}
\end{subequations}
Solving this is still NP-hard.
Therefore, we propose an equivalent problem in the following theorem.
\begin{theorem}
\label{RWMMSE}
With the optimum receive filters in \eqref{cerf_com} and \eqref{cerf_pri}, problem \eqref{lse_prob_lb} is equivalent to the following problem as
\begin{subequations}
\label{lse_prob_lb_eq}
\begin{align}
\label{lse_prob_lb_eq_obj}
&\min_\brP \quad f_2\slp \brP \srp \triangleq \sum_{k=1}^K \Tr \brW_{c,k} {\brM}_{c,k} + \brW_{p,k} {\brM}_{p,k} \srp \\
\label{lse_prob_lb_eq_con}
&\text{subject to} \quad \Tr \brP\brP^H \srp = \rho,
\end{align}
\end{subequations}
where
\begin{equation}
\label{weight_com}
\begin{aligned}
\brW_{c,k}=\mu_k \slp {\brM}_{c,k}^\text{MMSE} \srp^{-1}
\end{aligned}
\end{equation}
\begin{equation}
\label{weight_pri}
\begin{aligned}
\brW_{p,k}= \slp {\brM}_{p,k}^\text{MMSE} \srp^{-1}
\end{aligned}
\end{equation}
with $\mu_k \! = \! {\exp \slp \log \sll  {\brM}_{c,k}^\text{MMSE} \srl \srp}/{\sum_{j=1}^K \! \exp \slp \log \sll  {\brM}_{c,j}^\text{MMSE} \srl \srp}$.
\end{theorem}
\begin{IEEEproof}
\textit{See Appendix \ref{proof_RWMMSE}.}
\end{IEEEproof}
Due to the equivalence, a stationary point of problem \eqref{lse_prob_lb_eq} is also a stationary point of problem \eqref{lse_prob_lb}.
Therefore, both problems can be solved simultaneously.
However, jointly searching the precoding matrices while updating the optimum receive filters is a non-trivial task.
Thus, we adopt the BCD method in \cite{WMMSE11} to solve problem \eqref{lse_prob_lb_eq}.
In this approach, the optimum receive filters and weight matrices are first updated with the precoding matrices fixed, followed by updating the precoding matrices while keeping the receive filters and weight matrices fixed.
As the expressions for the optimum receive filters and weight matrices are explicitly given in \eqref{cerf_com}, \eqref{cerf_pri}, \eqref{weight_com}, and \eqref{weight_pri}, we thus focus on solving the precoding matrices in problem \eqref{lse_prob_lb_eq}.

Note that ${\brM}_{c,k}$ and ${\brM}_{p,k}$ in \eqref{lse_prob_lb_eq_obj} differ from \eqref{cemmse_com} and \eqref{cemmse_pri}, which assume given precoding matrices.
Problem \eqref{lse_prob_lb_eq} is difficult to solve because $\Pc$ and $\Pp$ are coupled in the objective function.
Besides, ${\brM}_{c,k}^\text{MMSE}$ depends on both $\Pc$ and $\Pp$, and ${\brM}_{p,k}^\text{MMSE}$ is a function of $\Pp$ only.
Consequently, the conventional Lagrange multiplier method is not applicable.
To address \eqref{lse_prob_lb_eq}, we adopt a new parameter $t \in \slp 0,1 \mrp$ to indicate the power allocated to the private signals.
When $t = 1$, the transmit power is fully assigned to the private signals.
By introducing this relaxation parameter, we can decouple the optimization of the common and private precoding matrices.
Thus, problem \eqref{lse_prob_lb_eq} can be rewritten as
\begin{subequations}
\label{lse_prob_lb_eq_t}
\begin{alignat}{2}
\label{lse_prob_lb_eq_t_obj}
&\min_{\Pc, \Pp, t} \quad && f_3\slp\Pc,\Pp, t \srp \triangleq  \sum_{k=1}^K \Tr \brW_{c,k} {\brM}_{c,k} \! + \! \brW_{p,k} {\brM}_{p,k} \srp \\
\label{lse_prob_lb_eq_t_con}
&\text{subject to} \quad && \Tr \Pc\Pc^H \srp = \rho\slp1-t\srp, \\
&&& \Tr \Pp\Pp^H \srp = \rho t.
\end{alignat}
\end{subequations}

Now, we can solve the above problem \eqref{lse_prob_lb_eq_t} by decomposing it into the following subproblems as
\begin{align}
\label{P1}
\slp \text{P1} \srp :&\min_{\Pp} \ g_1 \slp \Pp \srp \triangleq \sum_{k=1}^K \Tr \brW_{p,k} {\brM}_{p,k} \srp \\
&\text{subject to} \quad \Tr \Pp\Pp^H \srp = \rho t, \nonumber \\
\label{P2}
\slp \text{P2} \srp : &\min_{\Pc} \ g_2 \slp \Pc \srp \triangleq \sum_{k=1}^K \Tr \brW_{c,k} {\brM}_{c,k} \srp \\
&\text{subject to} \quad \Tr \Pc\Pc^H \srp = \rho\slp1-t\srp, \nonumber \\
\slp \text{P3} \srp : &\min_{t} \ g_3 \slp t \srp \triangleq \sum_{k=1}^K \Tr \brW_{c,k} {\brM}_{c,k} \! + \! \brW_{p,k} {\brM}_{p,k} \srp \\
&\text{subject to} \quad 0 < t \leq 1. \nonumber
\end{align}

The above decomposition follows from the fact that the common signal is first decoded by treating all private signals as interference.
Thus, the common precoding design should take the private precoding matrices into account.
In contrast, the private precoding design is not affected by the common precoding matrix.
Moreover, if $\Pc$ and $\Pp$ are fixed, $f_3\slp \Pc,\Pp,t \srp$ becomes solely a function of $t$.
As a result, during each iteration, we can first optimize $\Pp$ with fixed $t$.
The common precoding matrix $\Pc$ is then optimized by fixing $\Pp$ and $t$.
Finally, with obtained $\Pc$ and $\Pp$, we can determine the optimal $t$.
In the following, let us sequentially solve these subproblems.

\subsubsection{Solution of (P1)}
\label{solutionP1}
Let us denote $\whatH= [ \whatH_1,\cdots,\whatH_K ]$, $\brE = \mlp \brE_1, \cdots, \brE_K \mrp$, $\brW_z=\blkdiag\llp \brW_{z,1},\cdots,\brW_{z,K} \lrp$, and $\brD_z =\blkdiag\llp \brD_{z,1}^\text{MMSE},\cdots, \brD_{z,K}^\text{MMSE}\lrp$ for $z\in \llp c,p \lrp$ as the concatenations of the channels, the CSIT errors, the weight matrices, and the optimum receive filters, respectively.
The Lagrangian function of problem $\slp \text{P1} \srp$ is given by
\begin{equation*}
\begin{aligned}
L\slp \Pp,\lambda_1 \srp = \ & g_1 \slp \Pp \srp + \lambda_1 \slp \Tr \Pp\Pp^H \srp - \rho t \srp \\
= \ & \Tr \brW_p \srp - \Tr \Pp^H \brV \srp - \Tr \brV^H \Pp \srp \\
& + \Tr \brW_p \brD_p \whatH^H \Pp\Pp^H \whatH \brD_p^H \srp \\
& + \Tr \brOmega_{p}\brP_p\brP_p^H \srp + \No \Tr \brW_p\brD_p\brD_p^H \srp \\
& + \lambda_1 \Tr \Pp\Pp^H \srp - \lambda_1 \rho t, \\
\end{aligned}
\end{equation*}
where $\brV = \whatH\brD_p^H\brW_p$, $\brOmega_p = \diag\llp\brTheta \diagi\llp \brD_p^H\brW_p\brD_p \lrp \lrp$ with $\brTheta = \mlp \sigma_{e,1}^2 \one_{M\times N}, \cdots, \sigma_{e,K}^2 \one_{M\times N} \mrp$ being the error variance concatenation, and $\lambda_1$ is the Lagrange multiplier.
In $L\slp \Pp,\lambda_1 \srp$, the term associated with $\brOmega_{p}$ is determined by employing Lemma \ref{lemma_expectation} as
\begin{equation*}
\begin{aligned}
\Tr \brW_p \brD_p \bbE_{\brE} \mlp \brL_p \mrp \brD_p^H \srp = \Tr \brOmega_{p}\brP_p\brP_p^H \srp,
\end{aligned}
\end{equation*}
where $\brL_p = \brE^H \Pp\Pp^H \brE$.

Given $t^*$ computed from the last iteration, the unnormalized $\Pp$ can be calculated by applying the Lagrange multiplier method as
\begin{equation*}
\begin{aligned}
\overline{\brP}_p = \slp \brB + \lambda_1 \brI_M  \srp^{-1} \brV,
\end{aligned}
\end{equation*}
where $\brB = \whatH\brD_p^H\brW_p\brD_p\whatH^H + \brOmega_p$ and the Lagrange multiplier is $\lambda_1 = \frac{\No}{\rho t^*} \Tr \brW_p\brD_p\brD_p^H \srp$.
As a result, the solution to (P1) is given by
\begin{equation}
\label{optimalnorPp}
\begin{aligned}
\Pp = \sqrt{\rho t^*}\tildePp,
\end{aligned}
\end{equation}
where $\tildePp = \overline{\brP}_p/\sqrt{\Tr \overline{\brP}_p\overline{\brP}_p^H \srp}$ is the concatenation of normalized private precoding matrices.

\subsubsection{Solution to (P2)}
With the obtained $\Pp$, we derive the Lagrangian function of problem $\slp \text{P2} \srp$ as
\begin{equation*}
\begin{aligned}
L\slp \Pc,\lambda_2 \srp = \ & g_2 \slp \Pc \srp + \lambda_2 \slp \Tr \Pc\Pc^H \srp - \rho\slp1-t\srp \srp \\
= \ & \Tr \brW_c \srp - \Tr \Pc^H \brU \srp - \Tr \brU^H \Pc \srp \\
& + \Tr \brW_c \brD_c \whatH^H \brP\brP^H \whatH \brD_c^H \srp \\
& + \Tr \brOmega_{c}\brP\brP^H \srp + \No \Tr \brW_c\brD_c\brD_c^H \srp \\
& + \lambda_2 \Tr \Pc\Pc^H \srp - \lambda_2 \rho\slp1-t\srp, 
\end{aligned}
\end{equation*}
where we have $\brU = \sum_{k=1}^K \whatHk\slp\brD_{c,k}^\text{MMSE}\srp^H \brW_{c,k}$ and $\brOmega_{c} = \diag\llp\brTheta \diagi\llp \brD_c^H\brW_c\brD_c \lrp \lrp$.
Similarly, the term associated with $\brOmega_{c}$ is determined by employing Lemma \ref{lemma_expectation} as
\begin{equation*}
\begin{aligned}
\Tr \brW_c \brD_c \bbE_{\brE} \mlp \brL_c \mrp \brD_c^H \srp = \Tr \brOmega_{c}\brP\brP^H \srp,
\end{aligned}
\end{equation*}
where $\brL_c = \brE^H \brP\brP^H \brE$.

The unnormalized $\Pc$ can also be obtained using the Lagrange multiplier method as
\begin{equation*}
\begin{aligned}
\overline{\brP}_c = \slp \brA + \lambda_2 \brI_M  \srp^{-1} \brU,
\end{aligned}
\end{equation*}
where $\brA$ and the Lagrange multiplier are respectively given by
\begin{equation*}
\begin{aligned}
\brA = \whatH\brD_c^H\brW_c\brD_c\whatH^H + \brOmega_c,
\end{aligned}
\end{equation*}
\begin{equation*}
\begin{aligned}
\lambda_2 = \frac{1}{\rho \slp 1 - t^*\srp} \slp \No \Tr \brW_c\brD_c\brD_c^H \srp + \Tr \brA\Pp\Pp^H \srp \srp.
\end{aligned}
\end{equation*}
As a result, the solution to (P2) is
\begin{equation}
\label{optimalnorPc}
\begin{aligned}
\Pc = \sqrt{\rho \slp 1-t^*\srp}\tildePc,
\end{aligned}
\end{equation}
where $\tildePc = \overline{\brP}_c/\sqrt{\Tr \overline{\brP}_c\overline{\brP}_c^H \srp}$ is the normalized common precoding matrix.
After finding the solutions to (P1) and (P2), we should identify the power allocation between the common and private signals to maximize $f_3 \slp \Pc,\Pp,t\srp$.
In what follows, we solve (P3) to determine the optimal $t$ for each iteration.

\subsubsection{Solution to (P3)}
To obtain the optimal $t$, we first examine the derivative of $g_3\slp t \srp$ with respect to $t$ as
\begin{equation*}
\begin{aligned}
\frac{\partial g_3\slp t \srp}{\partial t} = & \sqrt{\frac{\rho}{1-t}} \Tr \brU^H\tildePc \srp  - \sqrt{\frac{\rho}{t}} \Tr \brV^H \tildePp \srp \\ 
& + \rho \Tr \slp \brA + \brB \srp \tildePp\tildePp^H \srp - \rho \Tr \brA \tildePc\tildePc^H \srp.
\end{aligned}
\end{equation*}
We find that $\lim_{t \rightarrow 0} \frac{\partial g_3\slp t \srp}{\partial t} = -\infty$ and $\lim_{t \rightarrow 1} \frac{\partial g_3\slp t \srp}{\partial t} = \infty$.
Thus, a root of $\frac{\partial g_3\slp t \srp}{\partial t}$ must exist within $\slp 0,1 \srp$.

Next, we compute the second-order derivative of $g_3\slp t\srp$ with respect to $t$ as
\begin{equation*}
\begin{aligned}
\frac{\partial^2 g_3\slp t \srp}{\partial t^2}  = \frac{\sqrt{\rho}}{2\slp 1-t\srp^{3/2}} \Tr \brU^H\widetilde{\brP}_c \srp + \frac{\rho}{2t^{3/2}} \Tr \brV^H \widetilde{\brP}_p \srp.
\end{aligned}
\end{equation*}
Since $t > 0$ and both $\brU^H\widetilde{\brP}_c$ and $\brV^H \widetilde{\brP}_p$ are positive definite, $\frac{\partial^2 g_3\slp t \srp}{\partial t^2}$ must be positive, implying that $\frac{\partial g_3\slp t \srp}{\partial t}$ is monotonically increasing.
Therefore, there is a unique $t$ within $\slp 0,1 \srp$. 
Since a closed-form solution for the root is not available, the optimal $t^\star$ can be obtained by employing the bisection method.
Finally, the precoding matrices are computed by 
\begin{equation}
\label{optimalPc}
\begin{aligned}
\Pc^\star = \sqrt{\rho\slp 1- t^\star \srp} \tildePc,
\end{aligned}
\end{equation}
\begin{equation}
\label{optimalPp}
\begin{aligned}
\Pp^\star = \sqrt{\rho t^\star} \tildePp.
\end{aligned}
\end{equation}
We establish an alternating algorithm by iteratively solving the constructed subproblems.
The procedures of the proposed robust precoding are summarized in Algorithm \ref{RS_RWMMSE}.

\begin{algorithm}  
\caption{Proposed Precoding Scheme}
\label{RS_RWMMSE}
\begin{algorithmic}
\STATE {Initialize $\brP=\brP^\text{init}$}
\REPEAT
\STATE {Compute $\brD_{p,k}^\text{MMSE}$ and $\brW_{p,k}$, $\forall k$ using \eqref{cerf_pri} and \eqref{weight_pri}, respectively.}
\STATE {Compute $\Pp$ using \eqref{optimalnorPp}.}
\STATE {Compute $\brD_{c,k}^\text{MMSE}$ and $\brW_{c,k}$, $\forall k$ using \eqref{cerf_com} and \eqref{weight_com}, respectively.}
\STATE {Compute $\Pc$ using \eqref{optimalnorPc}.}
\STATE {Compute $t^\star$ by applying bisection for $\frac{\partial g_3\slp t \srp}{\partial t} = 0$.}
\STATE {Compute $\Pc^\star$ and $\Pp^\star$ using \eqref{optimalPc} and \eqref{optimalPp}, respectively.}
\UNTIL{convergence}
\end{algorithmic}  
\end{algorithm}

\remark
The physical-layer implementation of the proposed precoding scheme can be achieved by following the same design as in \cite{RSWMMSE22} and \cite{RSMAfu22}.
Thus, we omit these aspects to avoid redundancy.
As will be shown in the simulation results, the proposed scheme achieves performance comparable to the WMMSE-SAA precoding scheme in \cite{RSWMMSE22}.
Both schemes share identical degrees of freedom.
Consequently, we also omit the detailed degree of freedom analysis.

\subsection{Convergence Analysis}
\label{ConverAnalysis}
To demonstrate the convergence, we will show that $f_3 \slp \Pc, \Pp, t \srp$ in \eqref{lse_prob_lb_eq_t_obj} is monotonically non-increasing in each iteration of Algorithm \ref{RS_RWMMSE}.
For the $i$-th iteration, with a fixed $t$, the minimization of problem (P1) with respect to $\Pp$ uniquely gives $f_3 \slp \Pc^{i-1},\Pp^{i}, t^{i-1} \srp$.
Thus, we have $f_3 \slp \Pc^{i-1},\Pp^{i}, t^{i-1} \srp \leq f_3 \slp \Pc^{i-1},\Pp^{i-1}, t^{i-1} \srp$.
Since $g_1 \slp \Pc \srp$ is block-wise convex with respect to $\Pc$, for given $\Pp$ and $t$, the minimization of problem $\slp \text{P2} \srp$ with respect to $\Pc$ also uniquely yields $f_3 \slp \Pc^i,\Pp^{i}, t^{i-1} \srp$, where $f_3 \slp \Pc^i,\Pp^{i}, t^{i-1} \srp \leq f_3 \slp \Pc^{i-1},\Pp^{i}, t^{i-1} \srp$.

Next, the optimal power allocation $t$ is uniquely determined by solving (P3), which leads to $f_3 \slp \Pc^i,\Pp^{i}, t^{i} \srp \leq f_3 \slp \Pc^{i},\Pp^{i}, t^{i-1} \srp$.
As a result, we can conclude that each iteration of the proposed algorithm is monotonically non-increasing since $f_3 \slp \Pc^i,\Pp^{i}, t^{i} \srp \leq f_3 \slp \Pc^{i-1},\Pp^{i-1}, t^{i-1} \srp$.
Due to the nature of the alternating minimization approach, the objective function $f_3 \slp \Pc, \Pp, t \srp$ decreases monotonically.

\subsection{Computational Complexity Analysis}
\label{CompComp}
Here, we analyze the computational complexity of the proposed precoding design and the conventional WMMSE-SAA scheme.
For simple analysis, we consider $M \geq NK$.
For the proposed precoding design, each iteration consists of the following steps.
The complexity of updating the receive filters \eqref{cerf_com} and \eqref{cerf_pri} is $\mathcal{O}\slp M^2 N K \srp$, while the update of the weight matrices \eqref{cemmse_com} and \eqref{cemmse_pri} requires the complexity of $\mathcal{O}\slp N^3 K \srp$.
The update of $t^\star$ is on the order of $\mathcal{O}\slp \log \frac{1}{\omega} \srp$, where $\omega$ is a required tolerance of the bisection method.
The computation of the precoding matrices \eqref{optimalPc} and \eqref{optimalPp} involves the matrix inversion operation, which has the complexity of $\mathcal{O}\slp M^3 \srp$ \cite{matcomp13}.

In constrast, for the WMMSE-SAA precoding scheme, the computational complexity of the receive filters and weight matrices are with $\mathcal{O}\slp M^2 N K L \srp$ and $\mathcal{O}\slp N^3 K L \srp$, respectively, where $L$ denotes the number of samples.
The update of the precoding matrices requires the optimization tools to solve a QCQP, which is on the order of $\mathcal{O}\slp \slp MN \slp K+1\srp \srp^{3.5} \srp$ \cite{QCQPcomp18}.
The above analysis shows that the proposed precoding design is more efficient than the WMMSE-SAA precoding scheme.
Moreover, since the proposed precoding design does not require memory to store the generated samples unlike the WMMSE-SAA, it is more suitable for the implementation in real-time communication systems.
For example, in the fully-loaded case with $M=NK$, the highest computational complexity of the proposed algorithm is $\mathcal{O}\slp M^3\srp$.
In contrast, the highest computational complexity of WMMSE-SAA is $\mathcal{O}\slp M^7\srp$.
This indicates that the proposed algorithm significantly reduces the computational complexity.

\section{Simulation Results}
\label{SimuResults}
In the numerical simulations, we evaluate the performance under the channel estimation error model with the noise variance of 1 assuming that all users have the same estimation error level $\sigma_e^2$.
We compare the proposed precoding design with the following conventional schemes: MRT, RBD \cite{BD04}, robust RBD (RRBD) \cite{GCI09}, SNS \cite{DLSNS22}, and WMMSE-SAA \cite{RSWMMSE22}.
Here, RWMMSE is obtained by deleting the common signal in the proposed precoding design.
The initial point of alternating RSMA algorithms is calculated by setting the common precoding matrix to the $M\times N$ submatrix of the left singular matrix of $\whatH$ with the power of $\rho\slp 1-t^\prime\srp$, where $t^\prime = \min\llp 1,\frac{1}{\rho \sigma_e^2}\lrp$, while the private precoding matrices are chosen as MRT with equal power allocation under the power of $\rho t^\prime$ \cite{RSWMMSE22}.

\begin{figure}[t]
\centering
\includegraphics[width=3.5in]{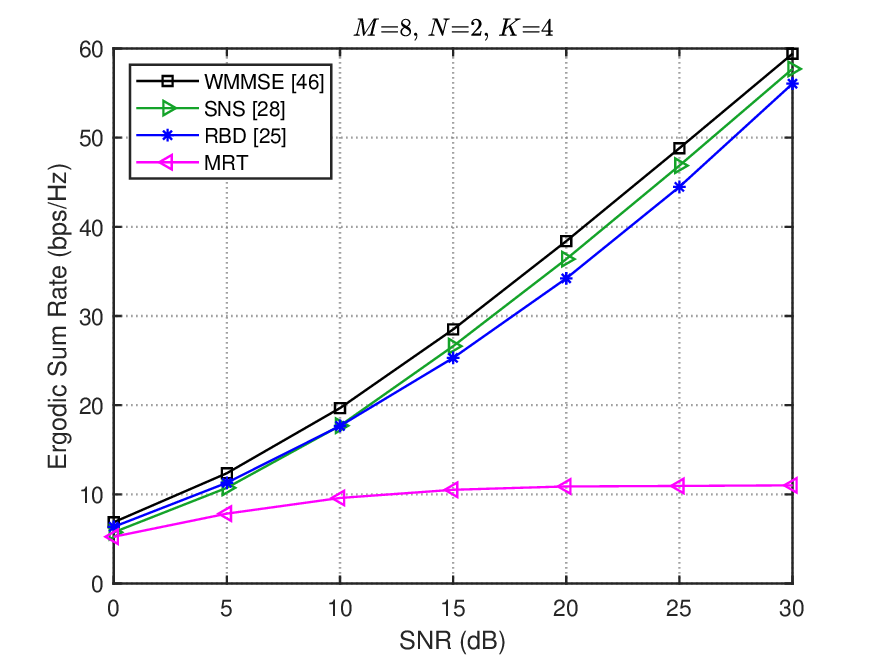}
\caption{Ergodic sum rate of MU-MIMO systems with $M=8$, $N=2$, and $K=4$ under perfect CSIT.}
\label{fig_SR_M8_N2_K4_PerfectCSI}
\end{figure}

First, we examine the ESR under perfect CSIT in Fig. \ref{fig_SR_M8_N2_K4_PerfectCSI}.
As $\sigma_e^2$ equals 0, RRBD boils down to RBD, and RWMMSE, WMMSE-SAA, and the proposed precoding design become equivalent to the conventional WMMSE precoding scheme \cite{WMMSE08}.
We can check that SNS performs worse than RBD at low SNR regions but displays better ESR than RBD at high SNR regions.
This arises from the fact that SNS does not consider the noise, thereby failing to improve the ESR at low SNR.
Among all schemes, WMMSE performs the best.

\begin{figure}[t]
\centering
\includegraphics[width=3.5in]{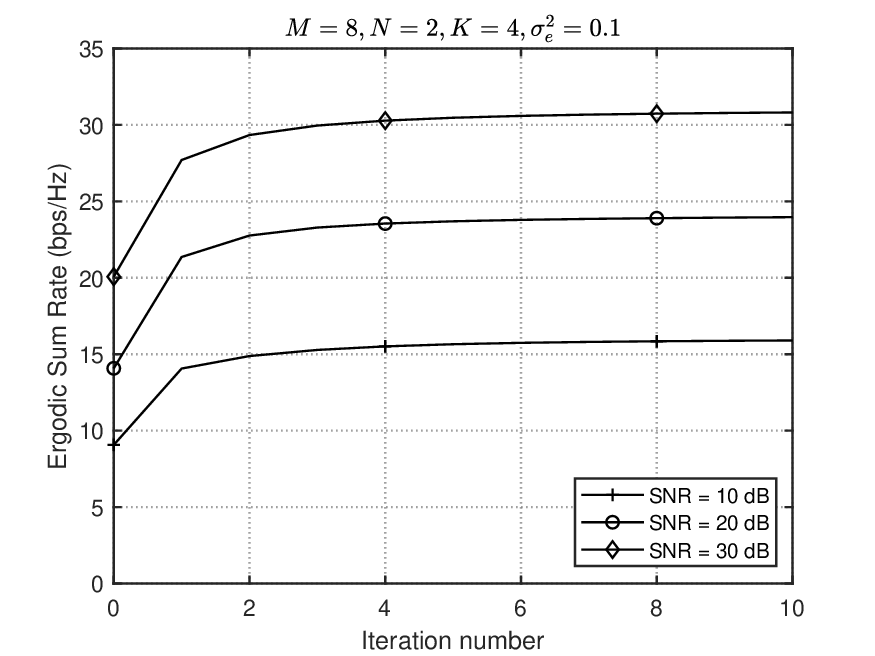}
\caption{Convergence of the proposed precoding designs of RSMA for MU-MIMO systems.}
\label{fig_convergence}
\end{figure}

Fig. \ref{fig_convergence} shows the convergence properties of the proposed precoding design.
We can see that the proposed scheme converges within 10 iterations.

\begin{figure}[t]
\centering
\includegraphics[width=3.5in]{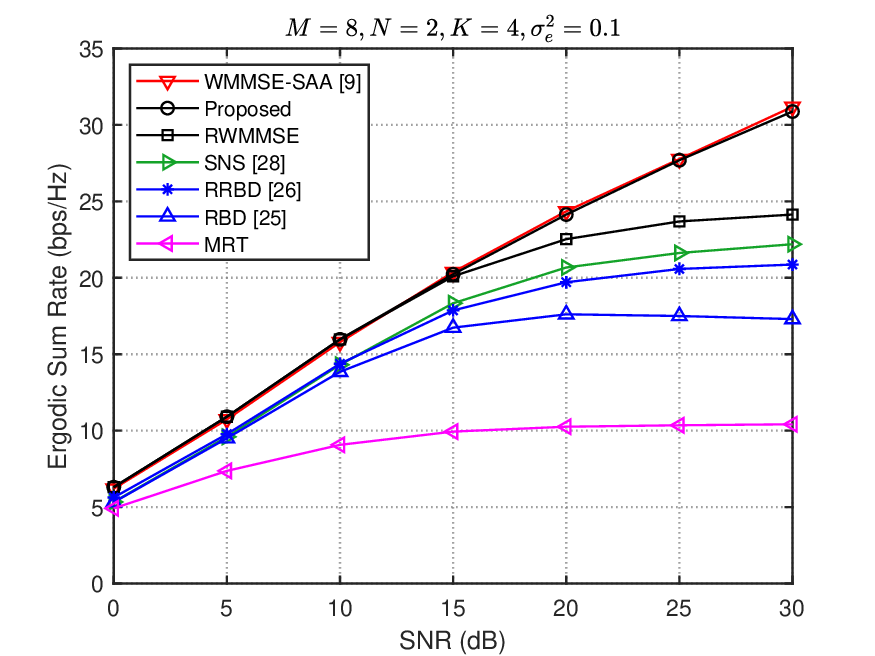}
\caption{Ergodic sum rate of MU-MIMO systems with imperfect CSIT.}
\label{fig_SR_M8_N2_K4}
\end{figure}

Next, we compare the ESR under different levels of channel estimation errors in Fig. \ref{fig_SR_M8_N2_K4} with $L = 1000$.
From the plot, non-RSMA schemes exhibit poor performance.
For example, the performance of RWMMSE saturates in high SNR.
In contrast, robust RSMA precoding schemes effectively mitigate the degradation due to estimation errors.
Among all precoding schemes, our proposed precoding design and the WMMSE-SAA precoding designs display the best ESR.
Thus, it can be verified that employing robust RSMA systems effectively manages interference caused by imperfect CSIT.
Also, it is important to note that the proposed scheme achieves almost the same performance as the conventional WMMSE-SAA.

\begin{figure}[t]
\centering
\includegraphics[width=3.5in]{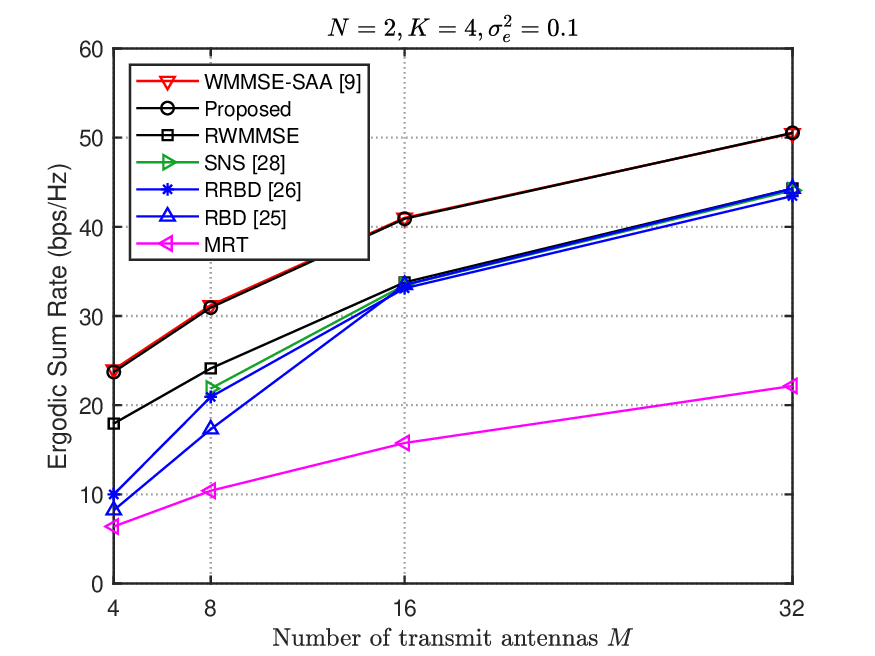}
\caption{Ergodic sum rate of MU-MIMO systems with different transmit antennas under SNR = 30 dB.}
\label{fig_SR_M481632_N2_K4}
\end{figure}

In Fig. \ref{fig_SR_M481632_N2_K4}, we illustrate the ESR for SNR = 30 dB with different numbers of transmit antennas.
In the case of $M \leq NK$, RWMMSE performs the best among non-RSMA schemes since it can allocate transmit power properly.
As the number of transmit antenna increases, the schemes with zero-forcing get close to RWMMSE due to the increased degree of freedom.
Again, it can be seen that our proposed scheme and WMMSE-SAA achieve the best in all system setups.

\begin{figure}[t]
\centering
\includegraphics[width=3.5in]{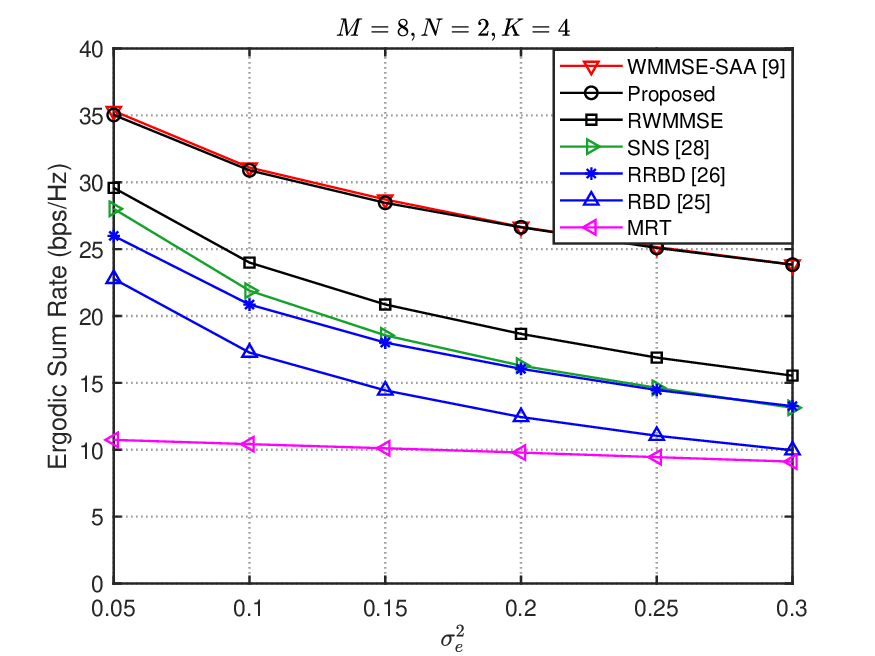}
\caption{Ergodic sum rate of MU-MIMO systems with different levels of estimation errors under SNR = 30 dB.}
\label{fig_SR_M8_N2_K4_VarE}
\end{figure}

To examine how the estimation error affects the system performance, we present the ESR with different levels of estimation errors in Fig. \ref{fig_SR_M8_N2_K4_VarE}.
It can be checked that all schemes experience a degraded ESR as $\sigma_e^2$ increases.
However, robust RSMA precoding schemes achieve a significant improvement even with a large $\sigma_e^2$ because of the adaptive power allocation among the common and private signals.
Thus, interference is reduced by employing RSMA in MU-MIMO systems when imperfect CSIT is available.

\begin{figure}[t]
\centering
\includegraphics[width=3.5in]{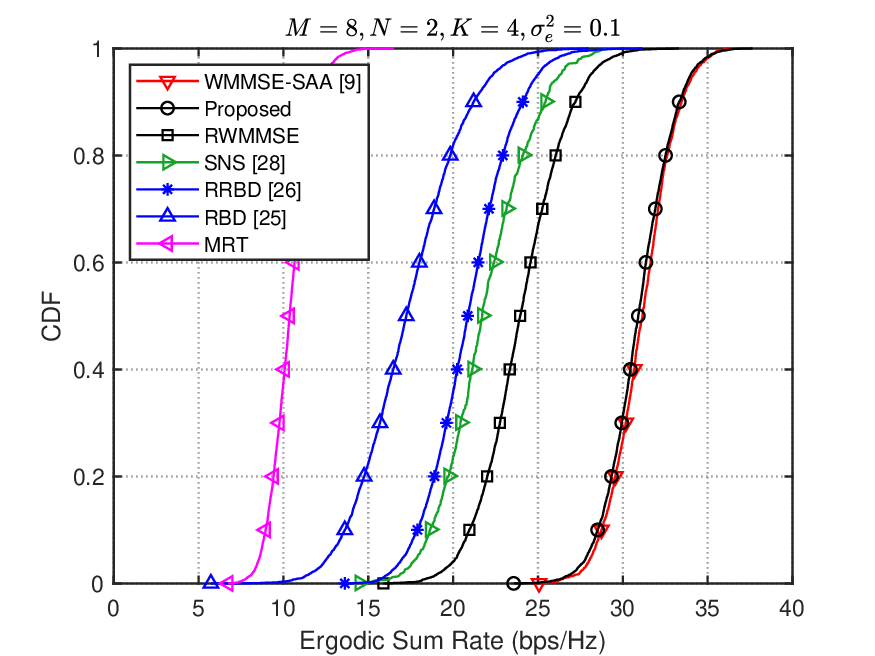}
\caption{CDF of MU-MIMO systems under SNR = 30 dB.}
\label{fig_CDF_M8_N2_K4}
\end{figure}

Next, we evaluate the robustness of different precoding schemes using the empirical cumulative distribution function (CDF).
Let $\hat{F}_N \slp x \srp$ denote the empirical CDF computed from the sample set $\llp x_i \lrp_{i=1}^N$ as
\begin{equation*}
\begin{aligned}
\hat{F}_N \slp x \srp = \frac{1}{N}\sum_{n=1}^N \mathcal{I}_{\llp x_n\leq x \lrp},
\end{aligned}
\end{equation*}
where $\mathcal{I}_{\llp x_n\leq x \lrp}$ is the indicator function, which equals 1 if $x_n \leq x$, and 0 otherwise.
By varying the value of $x$, we can empirically derive the corresponding CDF under the given constraints.
As shown in Fig. \ref{fig_CDF_M8_N2_K4}, for a given ESR, the value on the vertical axis represents the outage probability.
From the plot, it can be observed that most non-RSMA schemes fail to support transmission when the target rate is set to 30 bps/Hz.
In contrast, the proposed scheme and WMMSE-SAA achieve transmission at an outage probability of approximately 30\%.

\begin{figure}[t]
\centering
\includegraphics[width=3.5in]{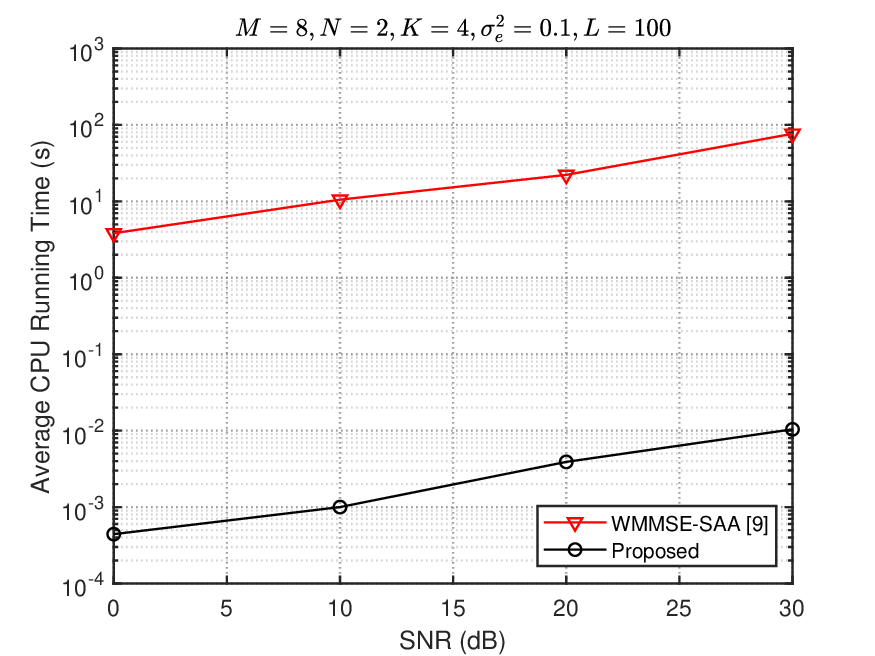}
\caption{Average CPU running time of RSMA for MU-MIMO systems with different SNRs.}
\label{fig_ACRTwithVarSNR}
\end{figure}

We then examine the complexity of different RSMA precoding schemes with SNR of 30 dB.
In addition to the worst-case computational complexity analysis in Section \ref{CompComp}, we evaluate the time complexity using MATLAB R2021a on a computer equipped with an Intel Core i7-11700 @2.5 GHz processor and 32 GB of RAM.
Fig. \ref{fig_ACRTwithVarSNR} presents the time complexity with different SNRs.
Increasing the transmit power has an significant impact on the time complexity.
From the plot, it is clear that the proposed scheme substantially reduce the complexity compared to the conventional method.

%

\section{Conclusion}
\label{Conclusion}
In this paper, we have proposed a novel precoding design to maximize the ESR in RSMA for MU-MIMO systems with imperfect CSIT.
To address the challenges associated with the non-smooth objective function, we have formulated a novel optimization problem using the generalized mutual information, and then transformed the formulated problem into a tractable one.
To solve the transformed problem, we have decomposed it into three subproblems.
Based on the sequential solutions of these subproblems, we have obtained an alternating precoding design by employing the BCD approach.
The simulation results have shown that our proposed precoding design achieves almost the same performance as the WMMSE-SAA scheme with much lower computational complexity.

\appendices

\section{Proof of Theorem \ref{RWMMSE}}
\label{proof_RWMMSE}
Since the optimization problems \eqref{lse_prob_lb} and \eqref{lse_prob_lb_eq} have the same constraints, we first focus on their objective functions.
The partial derivative of log-sum-exp is given by
\begin{equation*}
\begin{aligned}
\frac{\partial}{\partial x_k} \log \slp \sum_{k=1}^K \exp \slp x_k \srp \srp = \frac{\exp \slp x_k \srp}{\sum_{j=1}^K \exp \slp x_j \srp}.
\end{aligned}
\end{equation*}
Thus, the partial derivative of $f_1\slp \brP \srp$ with respect to $\Pc$ is calculated as
\begin{equation}
\label{f1_Pc}
\begin{aligned}
\nabla _{\Pc} & f_1 =  - \sum_{k=1}^K \mu_k \Big ( \whatHk \brF_k^{-1}\whatHk^H \Pc \slp {\brM}_{c,k}^\text{MMSE} \srp^{-1} \\
& + \whatHk \brF_k^{-1}\whatHk^H \Pc \slp {\brM}_{c,k}^\text{MMSE} \srp^{-1} \Pc^H\whatHk\brF_k^{-1} \whatHk^H \Pc \\
& + \bbE \mlp \Ek \brF_k^{-1}\whatHk^H \Pc \slp {\brM}_{c,k}^\text{MMSE} \srp^{-1} \Pc^H\whatHk\brF_k^{-1} \Ek^H \mrp \Pc \Big ),
\end{aligned}
\end{equation}
where $\mu_k = \exp \slp \log \sll {\brM}_{c,k}^\text{MMSE} \srl \srp/\sum_{j=1}^K \exp\slp \log \sll {\brM}_{c,j}^\text{MMSE} \srl \srp$.

Next, with the optimum receive filters in hand, the partial derivative of $f_2\slp \brP \srp$ with respect to $\Pc$ is calculated as
\begin{equation}
\label{f2_Pc}
\begin{aligned}
\nabla_{\Pc} f_2 \! = \! & - \sum_{k=1}^K \Big ( \whatHk  \brF_k^{-1}\whatHk^H \Pc \brW_{c,k} \\
& + \whatHk  \brF_k^{-1}\whatHk^H \Pc \brW_{c,k}  \Pc^H \whatHk\brF_k^{-1} \whatHk^H \Pc \\
& + \bbE\mlp \Ek  \brF_k^{-1}\whatHk^H \Pc \brW_{c,k} \Pc^H \whatHk\brF_k^{-1} \Ek^H \mrp \Pc \Big ).
\end{aligned}
\end{equation}
From \eqref{f1_Pc} and \eqref{f2_Pc}, it can be checked that they are identical when $\brW_{c,k}= \mu_k \slp {\brM}_{c,k}^\text{MMSE} \srp^{-1}$.

Similarly, we compute the partial derivatives of $f_1\slp \brP \srp$ and $f_2\slp \brP \srp$ with respect to the $k$-th private precoding $\Pk$ as
\begin{equation}
\label{f1_Pk}
\begin{aligned}
\nabla_{\Pk} & f_1 =  - \whatHk \brG_k^{-1}\whatH_k^H \Pk \slp {\brM}_{p,k}^\text{MMSE} \srp^{-1} \\
& - \sum_{j=1}^K \Big ( \mu_j \whatH_j \brG_j^{-1}\whatH_j^H \Pj \slp {\brM}_{c,j}^\text{MMSE} \srp^{-1}  \Pj^H \whatH_j \brG_j^{-1} \whatH_j^H\\
& +  \mu_j \bbE\mlp \brE_j \brG_j^{-1}\whatH_j^H \Pj \slp {\brM}_{c,j}^\text{MMSE} \srp^{-1}  \Pj^H \whatH_j \brG_j^{-1} \brE_j^H \mrp \\
& +  \whatH_j \brG_j^{-1}\whatH_j^H \Pj \slp {\brM}_{p,j}^\text{MMSE} \srp^{-1}  \Pj^H \whatH_j \brG_j^{-1} \whatH_j^H \\
& + \! \bbE \! \mlp \brE_j \brG_j^{-1}\whatH_j^H \Pj \! \slp {\brM}_{p,j}^\text{MMSE} \srp^{-1} \! \Pj^H \whatH_j \brG_j^{-1} \brE_j^H \mrp \! \Big ) \Pk,
\end{aligned}
\end{equation}
\begin{equation}
\label{f2_Pk}
\begin{aligned}
\nabla_{\Pk} & f_2 =  - \whatHk \brG_k^{-1}\whatH_k^H \Pk \brW_{p,k} \\
& - \sum_{j=1}^K \Big ( \whatH_j \brG_j^{-1}\whatH_j^H \Pj \brW_{c,j}  \Pj^H \whatH_j \brG_j^{-1} \whatH_j^H \\
& + \bbE\mlp \brE_j \brG_j^{-1}\whatH_j^H \Pj \brW_{c,j}  \Pj^H \whatH_j \brG_j^{-1} \brE_j^H \mrp \\
& + \whatH_j \brG_j^{-1}\whatH_j^H \Pj \brW_{p,j}  \Pj^H \whatH_j \brG_j^{-1} \whatH_j^H \\
& +\bbE\mlp \brE_j \brG_j^{-1}\whatH_j^H \Pj \brW_{p,j}  \Pj^H \whatH_j \brG_j^{-1} \brE_j^H \mrp \! \Big ) \Pk.
\end{aligned}
\end{equation}

Observing \eqref{f1_Pk} and \eqref{f2_Pk}, we can also notice that they are identical when $\brW_{c,k}= \mu_k \slp {\brM}_{c,k}^\text{MMSE} \srp^{-1}$ and $\brW_{p,k} = \slp {\brM}_{p,k}^\text{MMSE} \srp^{-1}$.
Now let us take the constraint into consideration.
If the partial derivatives are identical under the same equality constraint, the Karush-Kuhn-Tucker (KKT) conditions of problems \eqref{lse_prob_lb} and \eqref{lse_prob_lb_eq} are also the same \cite{CO04}.
Thus, a stationary point of problem \eqref{lse_prob_lb} is also a stationary point of problem \eqref{lse_prob_lb_eq}.
As a result, both problems can be solved at the same time.
This completes the proof.
$\hfill \blacksquare$

\ifCLASSOPTIONcaptionsoff
  \newpage
\fi



\begin{thebibliography}{1}

\bibitem{GPIP20}
J. Choi, N. Lee, S. N. Hong, and G. Caire, ``Joint user selection, power allocation, and precoding design with imperfect CSIT for multi-cell MU-MIMO downlink systems,'' \textit{IEEE Trans. Wireless Commun.}, vol. 19, no. 1, pp. 162--176, Jan. 2020.

\bibitem{ESLSMA13}
H. Yin, D. Gesbert, M. Filippou, and Y. Liu, ``A coordinated approach to channel estimation in large-scale multiple-antenna systems,'' \textit{IEEE J. Sel. Areas Commun.}, vol. 31, no. 2, pp. 264--273, Feb. 2013.

\bibitem{RDBPS19}
Z. Zhu, S. Huang, Z. Chu, F. Zhou, D. Zhang, and I. Lee, ``Robust designs of beamforming and power splitting for distributed antenna systems with wireless energy harvesting,'' \textit{IEEE Systems Journal}, vol. 13, no. 1, pp. 30--41, Mar. 2019.

\bibitem{LFBD08}
N. Ravindran and N. Jindal, ``Limited feedback-based block diagonalization for the MIMO broadcast channel,'' \textit{IEEE J. Sel. Areas Commun.}, vol. 26, no. 8, pp. 1473--1482, Oct. 2008.

\bibitem{CQIA12}
J.-S. Kim, S.-H. Moon, S.-R. Lee, and I. Lee, ``A new channel quantization strategy for MIMO interference alignment with limited feedback,'' \textit{IEEE Trans. Wireless Commun.}, vol. 11, no. 1, pp. 358--366, Jan. 2012.

\bibitem{RP23}
W. Zhou, D. Zhang, M. Debbah, and I. Lee, ``Robust precoding designs for multiuser MIMO systems with limited feedback,” \textit{IEEE Trans. Wireless Commun.}, vol. 23, no. 8, Aug. 2024.

\bibitem{SRRS16}
H. Joudeh and B. Clerckx, ``Sum-rate maximization for linearly precoded downlink multiuser MISO systems with partial CSIT: A rate-splitting approach,'' \textit{IEEE Trans. Commun.}, vol. 64, no. 11, pp. 4847--4861, Nov. 2016.

\bibitem{RTRS16}
H. Joudeh and B. Clerckx, ``Robust transmission in downlink multiuser MISO systems: A rate-splitting approach,'' \textit{IEEE Trans. Signal Process.}, vol. 64, no. 23, pp. 6227--6242, Dec. 2016.

\bibitem{RSWMMSE22}
A. Mishra, Y. Mao, O. Dizdar, and B. Clerckx, ``Rate-splitting multiple access for downlink multiuser MIMO: Precoding optimization and PHY-layer design,'' \textit{IEEE Trans. Commun.}, vol. 70, no. 2, pp. 874--890, Feb. 2022.

\bibitem{RSMIMO16}
B. Clerckx, H. Joudeh, C. Hao, M. Dai, and B. Rassouli, ``Rate splitting for MIMO wireless networks: A promising PHY-layer strategy for LTE evolution,'' \textit{IEEE Commun. Mag.}, vol. 54, no. 5, pp. 98--105, May 2016.

\bibitem{RSMAto23}
B. Clerckx et al., ``A primer on rate-splitting multiple access: Tutorial, myths, and frequently asked questions,'' \textit{IEEE J. Sel. Areas Commun.}, vol. 41, no. 5, pp. 1265--1308, May 2023.

\bibitem{HK81}
T. Han and K. Kobayashi, ``A new achievable rate region for the interference channel,'' \textit{IEEE Trans. Inf. Theory}, vol. 27, no. 1, pp. 49--60, Jan. 1981.

\bibitem{RSMA19}
Y. Mao, B. Clerckx, and V. O. K. Li, ``Rate-splitting for multiantenna non-orthogonal unicast and multicast transmission: Spectral and energy efficiency analysis,'' \textit{IEEE Trans. Commun.}, vol. 67, no. 12, pp. 8754--8770, Dec. 2019.

\bibitem{RSDPC20}
Y. Mao and B. Clerckx, ``Beyond dirty paper coding for multi-antenna broadcast channel with partial CSIT: A rate-splitting approach,'' \textit{IEEE Trans. Commun.}, vol. 68, no. 11, pp. 6775--6791, Nov. 2020.

\bibitem{RSMAfu22}
Y. Mao, O. Dizdar, B. Clerckx, R. Schober, P. Popovski, and H. V. Poor, ``Rate-splitting multiple access: Fundamentals, survey, and future research trends,'' \textit{IEEE Commun. Surveys Tuts.}, vol. 24, no. 4, pp. 2073--2126, 4th Quart., 2022.

\bibitem{RSMACRAN19}
D. Yu, J. Kim, and S. Park, ``An efficient rate-splitting multiple access scheme for the downlink of C-RAN systems,'' \textit{IEEE Wireless Commun. Lett.}, vol. 8, no. 6, pp. 1555--1558, Dec. 2019.

\bibitem{RSMAISAC21}
C. Xu, B. Clerckx, S. Chen, Y. Mao, and J. Zhang, ``Rate-splitting multiple access for multi-antenna joint radar and communications,'' \textit{IEEE J. Sel. Topics Signal Process.}, vol. 15, no. 6, pp. 1332--1347, Nov. 2021.

\bibitem{RSMAURLLC21}
O. Dizdar, Y. Mao, Y. Xu, P. Zhu, and B. Clerckx, ``Rate-splitting multiple access for enhanced URLLC and eMBB in 6G,'' in \textit{Proc. 17th Int. Symp. Wireless Commun. Syst. (ISWCS)}, Sep. 2021, pp. 1--6.

\bibitem{RSsate21}
L. Yin and B. Clerckx, ``Rate-splitting multiple access for multigroup multicast and multibeam satellite systems,'' \textit{IEEE Trans. Commun.}, vol. 69, no. 2, pp. 976--990, Feb. 2021.

\bibitem{RSMARIS22}
A. S. de Sena, P. H. J. Nardelli, D. B. da Costa, P. Popovski, and C. B. Papadias, ``Rate-splitting multiple access and its interplay with intelligent reflecting surfaces,'' \textit{IEEE Commun. Mag.}, vol. 60, no. 7, pp. 52--57, Jul. 2022.

\bibitem{RSMAUP23}
O. Abbasi and H. Yanikomeroglu, ``Transmission scheme, detection and power allocation for uplink user cooperation with NOMA and RSMA,'' \textit{IEEE Trans. Wireless Commun.}, vol. 22, no. 1, pp. 471--485, Jan. 2023.

\bibitem{RSmMIMO16}
M. Dai, B. Clerckx, D. Gesbert, and G. Caire, ``A rate splitting strategy for massive MIMO with imperfect CSIT,'' \textit{IEEE Trans. Wireless Commun.}, vol. 15, no. 7, pp. 4611--4624, Jul. 2016.

\bibitem{RSSOS23}
D. B. Amor, M. Joham, W. Utschick, ``Rate splitting in FDD massive MIMO systems based on the second order statistics of transmission channels,'' \textit{IEEE J. Sel. Areas Commun.}, vol. 41, no. 5, pp. 1351--1365, May, 2023.

\bibitem{RSMAIN23}
E. Sadeghabadi and S. D. Blostein, ``RSMA precoding design based on interference nulling and sum rate upper bound," \textit{IEEE Trans. Commun.}, vol. 71, no. 7, pp. 4091--4104, Jul. 2023.

\bibitem{BD04}
Q. H. Spencer, A. L. Swindlehurst, and M. Haardt, ``Zero-forcing methods for downlink spatial multiplexing in multiuser MIMO channels,'' \textit{IEEE Trans. Signal Process.}, vol. 52, no. 2, pp. 461--471, Feb. 2004.

\bibitem{GCI09}
H. Sung, S.-R. Lee, and I. Lee, ``Generalized channel inversion methods for multiuser MIMO systems,'' \textit{IEEE Trans. Commun.}, vol. 57, no. 11, pp. 3489--3499, Nov. 2009.

\bibitem{LPSC20}
A. R. Flores, R. C. de Lamare, and B. Clerckx, ``Linear precoding and stream combining for rate splitting in multiuser MIMO systems,'' \textit{IEEE Commun. Lett.}, vol. 24, no. 4, pp. 890--894, Apr. 2020.

\bibitem{DLSNS22}
A. Krishnamoorthy and R. Schober, ``Downlink MIMO-RSMA with successive null-space precoding,'' \textit{IEEE Trans. Wireless Commun.}, vol. 21, no. 11, pp. 9170--9185, Nov. 2022.

\bibitem{GSVD22}
L. Khamidullina, A. L. F. de Almeida, and M. Haardt, ``Multilinear generalized singular value decomposition (ML-GSVD) and its application to multiuser MIMO systems,'' \textit{IEEE Trans. Signal Process.}, vol. 70, pp. 2783--2797, May 2022.

\bibitem{CCCP20}
Z. Li, C. Ye, Y. Cui, S. Yang, and S. Shamai, ``Rate splitting for multi-antenna downlink: Precoder design and practical implementation,'' \textit{IEEE J. Sel. Areas Commun.}, vol. 38, no. 8, pp. 1910--1924, Aug. 2020.

\bibitem{OPBFRSMA23}
T. Fang and Y. Mao, ``Optimal beamforming structure for rate splitting multiple access,'' in \textit{Proc. IEEE Int. Conf. Acoust. Speech Signal Process. (ICASSP)}, Apr. 2024, pp. 9161--9165.

\bibitem{LSE23}
J. Park, J. Choi, N. Lee, W. Shin, and H. V. Poor, ``Rate-splitting multiple access for downlink MIMO: A generalized power iteration approach,'' \textit{IEEE Trans. Wireless Commun.}, vol. 22, no. 3, pp. 1588--1603, Sep. 2022.

\bibitem{NP99}
D. Bertsekas, \textit{Nonlinear Programming}, 2nd ed. Belmont, MA: Athena Scientific, 1999.

\bibitem{CO04}
S. Boyd and L. Vandenberghe, \textit{Convex Optimization}. Cambridge, U.K.: Cambridge Univ. Press, 2004.

\bibitem{CE06}
M. Biguesh and A. B. Gershman, ``Training-based MIMO channel estimation: A study of estimator tradeoffs and optimal training signals," \textit{IEEE Trans. Signal Process.}, vol. 54, no. 3, pp. 884--893, Mar. 2006.

\bibitem{LCRD22}
P. Singh, S. Tiwari, and R. Budhiraja, ``Low-complexity LMMSE receiver design for practical-pulse-shaped MIMO-OTFS systems," \textit{IEEE Trans. Commun.}, vol. 70, no. 12, pp. 8383--8399, Dec. 2022.

\bibitem{DLLF20}
J. Jang, H. Lee, S. Hwang, H. Ren, and I. Lee, ``Deep learning-based limited feedback designs for MIMO systems,'' \textit{IEEE Wireless Commun. Lett.}, vol. 9, no. 4, pp. 558--561, Apr. 2020.

\bibitem{DLLF22}
J. Jang, H. Lee, I.-M. Kim, and I. Lee, ``Deep learning for multi-user MIMO systems: Joint design of pilot, limited feedback, and precoding,'' \textit{IEEE Trans. Commun.}, vol. 70, no. 11, pp. 7279--7293, Nov. 2022.

\bibitem{JALFP22}
Z. Tang, L. Sun, D. Niyato, Y. Zhang, and A. Liu, ``Jammer-assisted secure precoding and feedback design for MIMO IoT networks,'' \textit{IEEE Internet Things J.}, vol. 9, no. 14, pp. 12241--12257, Jul. 2022.

\bibitem{GMI10}
G. Caire, N. Jindal, M. Kobayashi, and N. Ravindran, ``Multiuser MIMO achievable rates with downlink training and channel state feedback,'' \textit{IEEE Trans. Inf. Theory}, vol. 56, no. 6, pp. 2845--2866, Jun. 2010.

\bibitem{mmfairRSMA23}
B. Lee and W. Shin, ``Max-min fairness precoder design for rate-splitting multiple access: Impact of imperfect channel knowledge," \textit{IEEE Trans. Veh. Technol.}, vol. 72, no. 1, pp. 1355--1359, Jan. 2023.

\bibitem{logdet_tit01}
S. N. Diggavi and T. M. Cover, ``The worst additive noise under a covariance constraint," \textit{IEEE Trans. Inf. Theory}, vol. 47, no. 7, pp. 3072--3081, Nov. 2001.

\bibitem{WMMSE11}
Q. Shi, M. Razaviyayn, Z.-Q. Luo, and C. He, ``An iteratively weighted MMSE approach to distributed sum-utility maximization for a MIMO interfering broadcast channel," \textit{IEEE Trans. Signal Process.}, vol. 59, no. 9, pp. 4331–4340, Sep. 2011.

\bibitem{matcomp13}
G. H. Golub, C. F. V. Loan, \textit{Matrix Computations}, 4th ed. Baltimore, MD, USA: Johns Hopkins Univ. Press, 2013.

\bibitem{QCQPcomp18}
P. Patil, B. Dai, and W. Yu, ``Hybrid data-sharing and compression strategy for downlink cloud radio access network,'' \textit{IEEE Trans. Commun.}, vol. 66, no. 11, pp. 5370--5384, Nov. 2018.

\bibitem{WMMSE08}
S. S. Christensen, R. Agarwal, E. De Carvalho, and J. M. Cioffi, ``Weighted sum-rate maximization using weighted MMSE for MIMO-BC beamforming design,'' \textit{IEEE Trans. Wireless Commun.}, vol. 7, no. 12, pp. 4792–4799, Dec. 2008.

\end{thebibliography}
\end{document}